\documentclass[12pt]{iopart}
\usepackage{epsfig}
\usepackage{amssymb}
\usepackage{graphicx}
\usepackage{version}

\begin{document}

\title[Coarsening in the $3d$ RFIM]{Scaling and super-universality in the coarsening
dynamics of the $3d$ random field Ising model}
\vskip 10pt
\author{Camille Aron}
\address{
Universit\'e Pierre et Marie Curie - Paris VI, 
Laboratoire de Physique Th\'eorique et Hautes \'Energies,
4 Place Jussieu,
75252 Paris Cedex 05 France}
\author{Claudio Chamon}
\address{Physics Department, Boston University, 
Boston, MA 02215, USA}
\author{Leticia F. Cugliandolo and Marco Picco}
\address{
Universit\'e Pierre et Marie Curie - Paris VI, 
Laboratoire de Physique Th\'eorique et Hautes \'Energies,
4 Place Jussieu,
75252 Paris Cedex 05 France}

\vskip 10pt

\begin{abstract}
  We study the coarsening dynamics of the three-dimensional random
  field Ising model using Monte Carlo numerical simulations.  We test
  the dynamic scaling and super-scaling properties of global and
  local two-time observables.  We treat in parallel the
  three-dimensional Edward-Anderson spin-glass and we recall results
  on Lennard-Jones mixtures and colloidal suspensions to highlight the
  common and different out of equilibrium properties of these glassy
  systems.
\end{abstract}
\date{today}
%\maketitle

\section{Introduction}

The physics of domain growth is well understood~\cite{Alan,Puri}.
Just after the initial thermal quench into the ordered phase, the
spins in a ferromagnetic system 
tend to order and form domains of the equilibrium states.  In
clean systems the ordering dynamics is governed by the symmetry and
conservation properties of the order parameter. When impurities are
present the dynamics are naturally slowed down by domain-wall
pinning~\cite{Nattermann}.  The dynamic scaling hypothesis
%that is well verified experimentally and numerically,
states that the time-dependence in any macroscopic observable enters
only through a growing length scale, $R(t)$, either the instantaneous
{\it averaged} or {\it typical} domain radius.  However, a complete
description of the phenomenon is lacking. In the pure cases the
scaling functions are not known analytically and no fully satisfactory
approximation scheme to estimate them is known~\cite{Alan}. In
presence of disorder the limitations are more severe in the sense that
the growth laws are derived by assuming that the relaxation is driven
by activation over free-energy barriers and the properties of the
latter are estimated with energy balancing arguments applied to single
interfaces that are hard to put to the test. Even in the relatively
simple random bond Ising model (RBIM)  the time dependence of the
growth law remains a subject of controversy~\cite{Heiko}.

Quenched randomness may be weak or strong in the sense that the first
type does not change the nature of the low-temperature phase, as in
the random bond or random field Ising (RFIM) models, or it can change it as
in spin-glasses. Fisher and Huse conjectured that in the first class
of systems, once the scaling hypothesis is used 
to describe the long times dynamics, so that times and
lengths are measured in units of $R(t)$, no out of equilibrium
observable depends on the quenched
randomness~\cite{Fisher-Huse} and their scaling functions are thus 
identical to the ones 
of the pure limit. This is the so-called `{\it
  super-universality}' hypothesis in coarsening phenomena.  Tests of
this hypothesis as applied to the equal-times two-point function of
the $3d$ RFIM and the $2d$ RBIM appeared
in~\cite{Rao} and \cite{Puri-Parekh}, respectively, and the
distribution of domain areas in the $2d$ RBIM in \cite{Alberto3}.

The dynamics of generic glassy systems is less well understood but
presents some similar aspects to those mentioned above.  The droplet
model of finite-dimensional spin-glasses is based on the assumption
that in the low-temperature phase these systems also undergo domain
growth of two competing equilibrium states~\cite{Fisher-Huse}. In the
mean-field limit spin-glasses have, though, a very different kind of
dynamics~\cite{Cuku94,Bacukupa} that cannot be associated to a simple
growth of two types of domains. Numerical studies of the $3d$
Edwards-Anderson (EA)
model~\cite{3dEA-num,3dEA-num2,3dEA-num3,Ludovic,Victor} have not been
conclusive in deciding for one or the other type of evolution and, in
a sense, show aspects of both. A one-time dependent
`coherence'-length, $R(t)$, has been extracted from the distance and
time dependence of the {\it equal-time} overlap between two replicas
evolving independently with the same quenched disordered
interactions~\cite{3dEA-num2,3dEA-num3,Victor}. A power-law $R(t)\sim
t^{1/z(T)}$ with the dynamic exponent $z(t)=z(T_c) T_c/T$ fits the
available data for the $3d$ EA and $z(T_c) = 6.86(16)$ with
Gaussian~\cite{Victor} and $z(T_c) = 6.54(20)$ with
bimodal~\cite{3dEA-num2,3dEA-num3} couplings. Still, it was claimed in~\cite{Victor} that the overlap decays to zero as a power law at long distances and long times such that $r/R(t)$ is fixed, implying that there are more than two types of growing domains in the low temperature phase.

A two-time dependent length, $\xi(t,t_w)$, can be extracted from the analysis of the
spatial decay of the correlation between two spins in the same system
at distance $r$ and different times $t$ and $t_w$ after
preparation~\cite{Castillo}.  The latter method is somehow more
powerful than the former one in the sense that it can be easily
applied to glassy problems without quenched disorder.  If there is
only one characteristic length-scale in the dynamics $R(t)$ should be
recovered as a limit of $\xi(t,t_w)$ but this fact has not been
demonstrated.

The mechanism leading to the slow relaxation of structural glasses is
also not understood. Still, molecular dynamic studies of Lennard-Jones
mixtures~\cite{Castillo-Parsaeian} and the analysis of confocal
microscopy data in colloidal suspensions~\cite{Weeks} show that
two-time observables have similar time dependence as in the $3d$ EA
model. Two-time correlations scale using ratios of one-time growing
functions that, however, cannot be associated to a domain radius
yet. A two-time correlation length $\xi$ with characteristics similar
to the one in the $3d$ EA can also be defined and measured.

The understanding of dynamic fluctuations in out of equilibrium
relaxing systems appears as a clear
challenge~\cite{Chamon-Cugliandolo}.  In systems with quenched
randomness different sample regions feel a different environment and
one expects to see their effect manifest in different ways working at
fixed randomness.  The effect of quenched randomness is at the root of
Griffiths singularities in the statics of disordered systems, for
instance. In structural or polymer glasses there are no quenched
interactions instead, but still one expects to see important
fluctuations in their dynamic behaviour both in metastable equilibrium
and in the glassy low temperature regime. The question of whether the
fluctuations in generic glassy systems resemble those in coarsening
systems has only been studied in a few solvable cases such as the
model of ferromagnetic coarsening in the large $N$ limit~\cite{Chcuyo}
and the Ising chain~\cite{Peter}.

In this paper we study ferromagnetic ordering
in the $3d$ RFIM following a quench from infinite temperature and we
compare it to the dynamics of the $3d$ EA spin-glass and particle 
glassy systems.  Our aim is to
signal which aspects of their out of equilibrium evolution differ and
which are similar by focusing on freely relaxing observables -- no
external perturbation is applied to measure linear responses.  We test
the scaling and super-universality hypothesis in the RFIM and we
explicitly show that the latter does not apply to the EA
model.  We analyse the spatio-temporal fluctuations in the coarsening
problem and we compare them to the ones found in
spin-glasses~\cite{Ludovic,Castillo}, the $O(N)$ ferromagnetic coarsening
in the large $N$ limit~\cite{Chcuyo}, and other glassy
systems~\cite{Weeks,Sellitto,Parsaeian-Castillo}.

The organisation of the paper is the following. In
Sect.~\ref{sec:model} we define the models and we describe the
numerical procedure.  Section~\ref{sec:growing-length} is devoted to
the study of the growing length scale, $R$, the scaling and
super-universality hypothesis, and the two-time growing length, $\xi$.
In Sect.~\ref{sec:fluctuations} we focus on the local fluctuations of
two time observables. We study two-time coarse-grained
correlations and we analyse their statistical properties as time
evolves. Finally, in Sect.~\ref{sec:conclusions} we present our
conclusions.

\section{The models}
\label{sec:model}

Two varieties of quenched disorder are encountered in spin models:
randomness in the strength of an externally applied magnetic field and
randomness in the strength of the bonds. The RFIM and the EA
spin-glass are two archetypal examples of these. In this Section we
present their definitions and we recall some of their main properties. 
%We shall try
%to put into perspective the results for the RFIM with a parallel
%treatment of the EA model and a summary of measurements in Lennard-Jones 
%mixtures and colloidal suspensions.

\subsection{The Random Field Ising Model}

The $3d$ Random Field Ising model (RFIM) is defined by the
Hamiltonian~\cite{Imry-Ma}
\begin{equation}
 	H = - J \sum_{\langle i,j\rangle} s_i s_j - \sum_{i} H_i s_i \,.
\end{equation}
The first term encodes short range ferromagnetic ($J>0$) interactions
between nearest neighbour Ising spins, $s_i=\pm 1$, placed on the
nodes of a cubic lattice with linear size $L$.  $H_i$ represents a
local random magnetic field on site $i$. We adopt a bimodal
distribution for these independent identically distributed random
variables ($H_{i} = \pm H$ with equal probability). $H$ quantifies the
strength of the quenched disorder.  Hereafter we set $J=1$ and we use
units in which $k_B=1$.

The RFIM is relevant to a large class of materials due to the presence
of defects that cause random fields. Dilute anisotropic
antiferromagnets in a uniform field are the most studied systems
expected to be described by the RFIM.  Several review articles
describe its static and dynamic behaviour~\cite{Nattermann} and the
experimental measurements in random field samples have been summarized
in~\cite{Belanger}. Dipolar glasses also show aspects of random field
systems~\cite{dipolar}.

In the case $H = 0$, the RFIM reduces to the well known clean Ising
model with a phase transition from a paramagnetic to a ferromagnetic
state occurring at $T_c \simeq 4.515$.  It is well established that in
$d=3$ (not in $d=2$) there is a phase separating line on the $(T,H)$
plane joining $(T_c,H=0)$ and $(T=0,H_c)$.
%A sufficiently strong magnetic field destroys all possible long range. 
At $T=0$ and small magnetic field, it has been rigorously proven that
the state is ferromagnetic~\cite{Imbrie,Bricmont}.  The nature of the
transition close to zero temperature has been the subject of some
debate. Claims of it being first order~\cite{Young} have now been
falsified and a second order phase transition has been 
proven~\cite{Rieger-RFIM,Middleton}.  It was also argued but not established
that there might be a spin-glass phase close to
$(T=0,H_c)$~\cite{Mezard}.  Recent numerical simulations yield
$H_c\simeq 2.215(35)$ at $T=0$~\cite{Swift,Malakis}.
% $H_c\sim 2.28$ for Gaussian distributed random fields~\cite{Machta}. 

\subsection{The Edwards-Anderson spin-glass}

The $3d$ 
Edwards-Anderson (EA) spin glass is defined by 
\begin{equation}\label{eq:EA}
	H =  - \sum_{\langle i,j\rangle} J_{ij} s_i s_j
\,.
\end{equation}
The interaction strengths $J_{ij}$ act on nearest neighbours on a
cubic three-dimensional lattice and are independent identically
distributed random variables. We adopt a bimodal distribution, 
$J_{ij} = \pm 1$ with equal probability. This 
model undergoes a static phase transition from a paramagnetic to a 
spin-glass phase at $T_g\simeq 1.14(1)$~\cite{Ballesteros}. The nature of the 
low temperature static phase is not clear yet and, as for the 
out of equilibrium relaxation, two pictures developed around
a situation with only two equilibrium states as proposed in the 
droplet model and a much more complicated vision emerging from 
the solution of the Sherrington-Kirkpatrick model, its 
mean-field extension~\cite{Fischer-Hertz}. 

\subsection{Numerical methods}

We focus here on the out of equilibrium relaxation in the ordered
phase. We simulate the dynamics following an instantaneous quench from
infinite temperature at the initial time, $t=0$, by choosing a random
initial condition: $s_i(t=0)=\pm 1$ with probability one half. The
order parameter is not conserved during the evolution.  We use the
continuous time Monte Carlo (MC) procedure~\cite{Bortz,Dall,Novotny}.
This algorithm, which is nothing else than a re-organisation of the
standard Monte Carlo rule, is rejection free. This makes it
spectacularly faster than standard MC which would have a rejection
rate close to $1$ in the ferromagnetic phase of the RFIM. Times are
expressed in usual Monte Carlo steps (MCs): $1$ MCs corresponds to
$N=L^d$ spin updates with the standard MC algorithm. The way to
translate from the continuous time MC to standard MC units, in which 
we present our results, is
explained in \cite{Bortz,Dall,Novotny}.

We study the relaxation dynamics with non-conserved order parameter 
in the ($d=3$) ferromagnetic phase of
the RFIM at relatively low temperature and small applied field.
Interesting times are not too short -- to avoid a short transient
regime -- and not too long -- to avoid reaching equilibration (in
ferromagnetic coarsening a non-zero magnetization density indicates
that the coarsening regime is finished and other more refined methods
are used in the spin-glass case~\cite{comment-refined}). 
We delay equilibration by taking
large systems since the equilibration time rapidly grows with the size
of the lattice.  A reasonable numerical time-window is $[10^3, 10^7]$
MCs.  We show results obtained using lattices with $L = 250$ ($N =
1.5\times 10^7$ spins) in the RFIM and $L=100$ ($N=10^6$ spins) in the
spin-glass.  We checked that finite size effects are not important in
any of these cases for averaged quantities.

\section{The typical growing length}
\label{sec:growing-length}

In this Section we study the typical growing length (a geometric
object) in the RFIM and the EA model. We establish scaling and
super-universality relations for three types of correlations functions
(statistical objects). Two of them involve
either two space points and one time, or one space point and two times,
and are the usual observables studied in coarsening phenomena.  The
third one is commonly used in the study of glassy systems where
two-point correlations are not sufficient to characterize the dynamics
of the systems~\cite{Ludovic,Castillo,Castillo-Parsaeian,Weeks} 
and allows for the definition 
of a two-time dependent length that we can compare to the one 
obtained in the $3d$ EA model and glassy particle systems.

\subsection{The RFIM}

During the ferromagnetic coarsening regime, there are as many positive
as negative spins in such a way that the magnetization density stays zero in
the thermodynamic limit and weakly fluctuates around zero for finite
size systems.  Everywhere in the sample, there is a local competition
between growing domains. Eventually, after an equilibration time
$\tau_{eq}$ (that diverges with the system size), one of the two
phases conquers the whole system scale.

In the coarsening regime (times shorter than $\tau_{eq}$) dynamic
scaling~\cite{Alan} applies and the growth of order is characterized
by a {\it typical domain radius}, $R(t;T,H)$, that increases in time
and depends on the control parameters, $T$ and $H$, and the dimension
of space, $d$~\cite{Alberto1}.  While in the absence of impurities it
is clearly established that, for non-conserved order parameter dynamics, 
the domain length $R$ grows as $R\sim
t^{1/2}$ independently of $d$~\cite{Alan} with a prefactor that
monotonically decreases upon increasing temperature~\cite{Alberto2},
the functional form of $R$ is less clear in random cases. Scaling
arguments based on the energetics of single
interfaces~\cite{Nattermann,Villain,Grinstein,Fernandez} predict a
crossover from the pure case result at short time-scales when it is
easy to inflate, to a logarithmic growth,
\begin{equation} \label{eq:R_scaling}
	R(t;H,T) = 
\frac{T}{H^2} \ln\left(t/\tau(T,H)\right)
\; . 
\end{equation}
The fact that the prefactor grows with $T$ (as opposed to what happens
for pure curvature driven dynamics~\cite{Alberto2}) is due to the
activated character of the dynamics. Several proposals for the
characteristic time $\tau$ exist: $\tau \sim
(T/H^2)^2$~\cite{Villain,Puri} and $\tau \sim \tau_0 e^{A(T)/H^2}$
with $A(T)$ a weakly temperature dependent function~\cite{Rao}.  To ease
the notation in what follows we do not write explicitly the $T$ and
$H$ dependence of $R$.

From the point of view of the renormalization group, all points within the
ferromagnetic region of the $(T,H)$ phase diagram flow to the stable,
zero-temperature, zero-disorder sink. Hence, randomness and
temperature should be irrelevant in equilibrium at $T<T_c$. The
super-universality hypothesis states that for non equilibrium ordering
dynamics, once lengths are scaled with the typical length $R$,
quenched random fields are irrelevant and all scaling functions are
the ones of the pure $3d$ Ising system at $T=0$ with non-conserved order
parameter.

\subsubsection{The equal-time spatial correlation.}

\begin{figure}
\begin{flushright}
 \includegraphics[width=4.75cm,angle=-90]{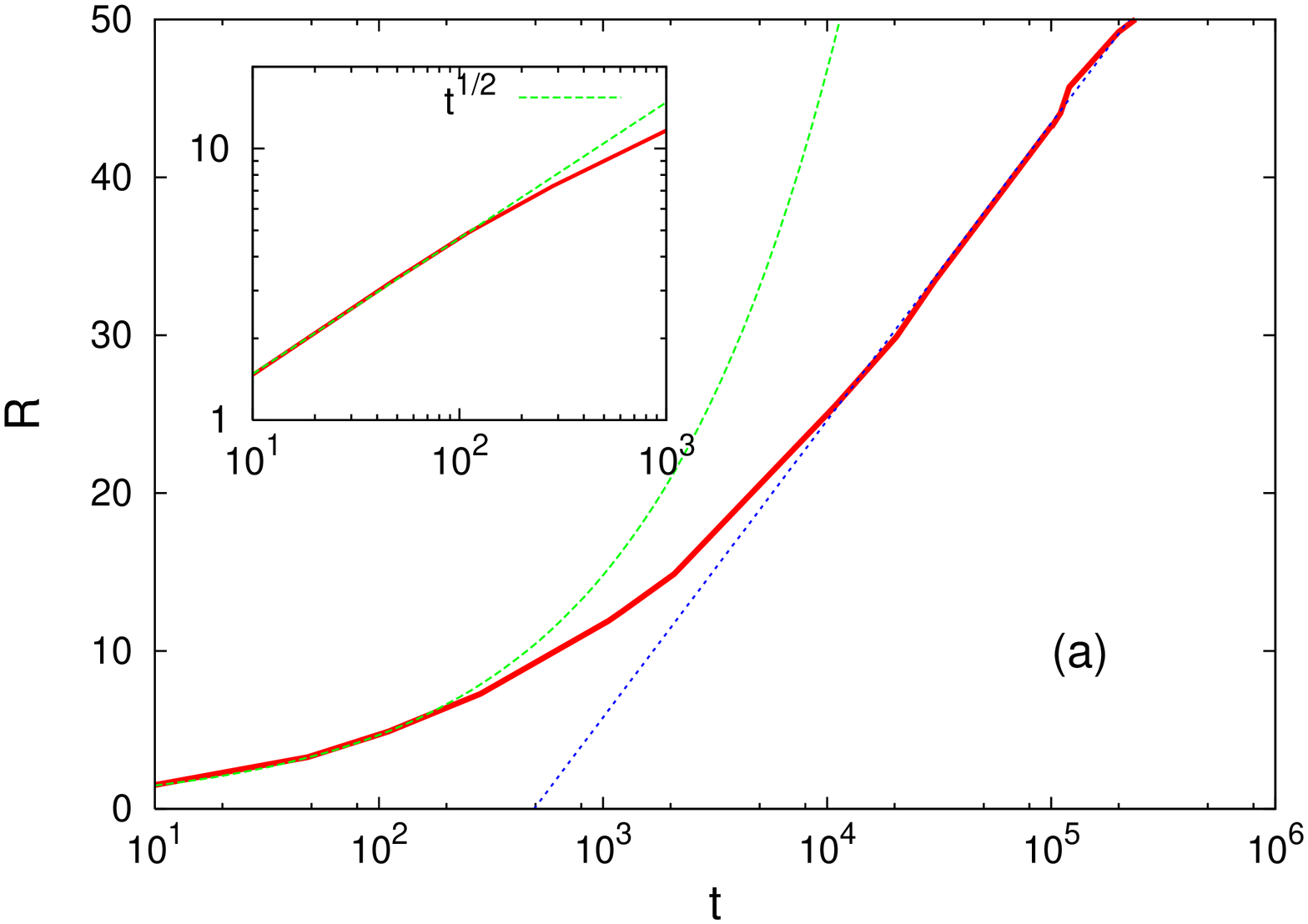}
 \includegraphics[width=4.75cm,angle=-90]{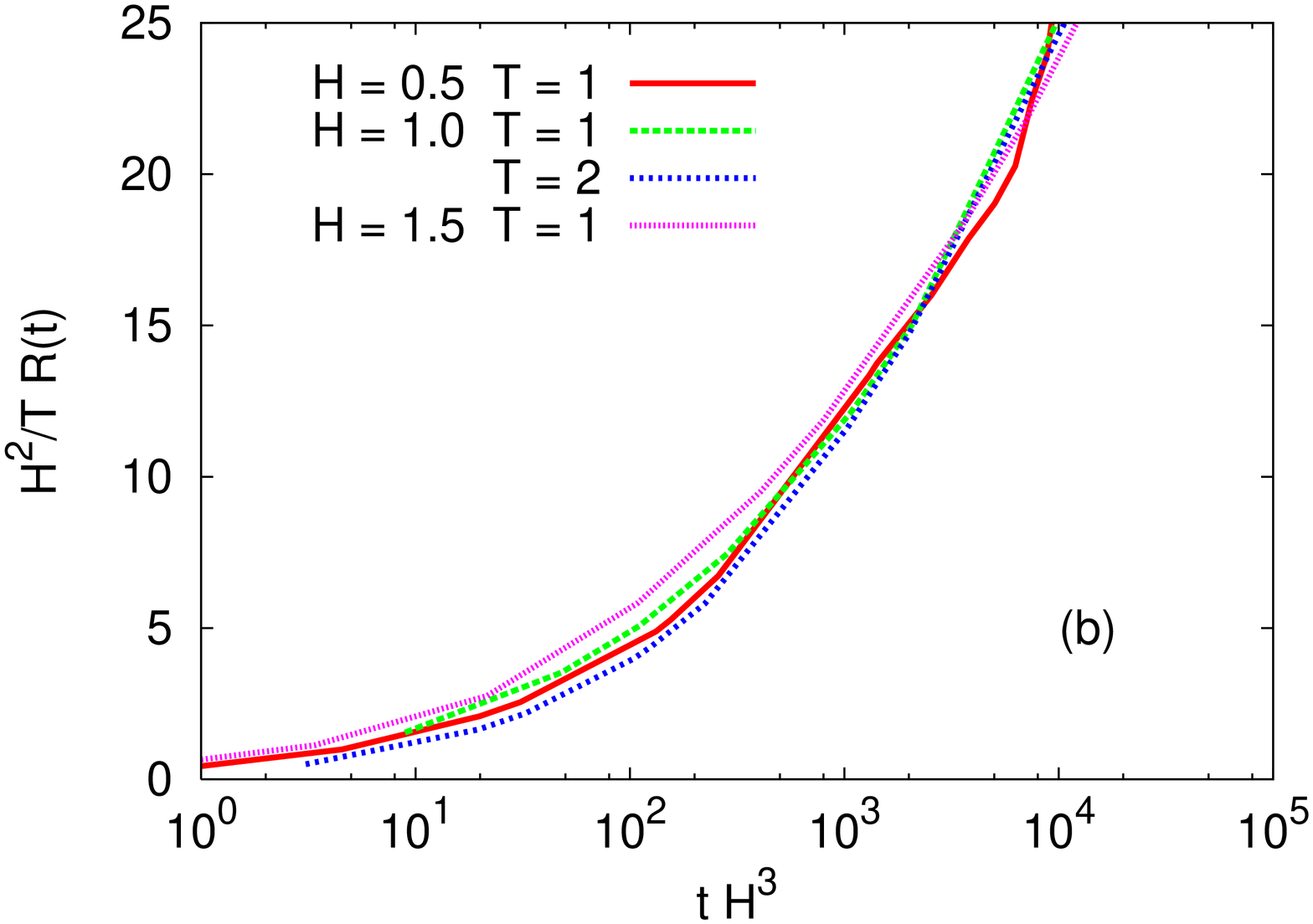}
\end{flushright}
\begin{center}
  \caption{\footnotesize \label{fig:RF_R} (a)~With line-points (red),
    the growing length $R(t)$ at $T = 1$ and $H = 1$. The green curve
    is the power law $\sqrt{t}$ that describes well the data at short
    times, right after the temperature quench. The blue line is a
    logarithmic law apt to describe the behaviour at longer
    time-scales.  In the inset: the same data in a log-log scale to
    highlight the quality of the $\sqrt{t}$ behaviour at short times.
    (b)~Study of the dependence of $R$ on the parameters $T$ and $H$
    for two values of $T$ and three random field strengths $H$ given
    in the key.}
\end{center}
\end{figure}

A careful analysis of the field and time dependence of the growing
length scale together with tests of the scaling hypothesis applied to
the equal-time correlation
\begin{equation}
  C_2(r;t) \equiv \langle s_i(t) s_j(t)\rangle _{\vert \vec r_i - \vec r_j
\vert = r}\; ,
\end{equation}
where the average runs over all spins in the sample, 
appeared in~\cite{Rao,Oguz}. 
%(The angular brackets indicate an
%average over the whole sample and/or thermal noise.) 
 In the
coarsening regime, at distances $a \ll r \ll L$ with $a$ the lattice 
spacing and $r/R(t)$ finite,
$C_2(r;t)$ is expected to depend on $r$ and time $t$ only through the
ratio $r/R$, 
\begin{equation}
C_2(r;t) \simeq m_{eq}^2 \; f_2(r/R(t))
\; , 
\label{eq:C2-scaling}
\end{equation} 
with $m_{eq}$ the equilibrium magnetization density (that decreases
with increasing $T$ and/or $H$), $\lim_{x\to 0} f_2(x) = 1$ and
$\lim_{x\to \infty} f_2(x) =0$. Since the spatial decay is approximately exponential, $C_2(r;t) \propto
e^{-r/R(t)}$ for not too long $r$, we use this functional form to
extract $R$ from the data fit at each set of parameters $(T,H,t)$.  Figure~\ref{fig:RF_R}~(a) shows that the growing length $R$ has two regimes: shortly after the quench $R$ grows as $t^{1/2}$
like in the pure case and it later crosses over to a logarithmic
growth. This is consistent with previous numerical studies in
$2d$~\cite{Puri-Parekh,Scott} and $3d$~systems~\cite{Rao,Oguz}.  In
Fig.~\ref{fig:RF_R}~(b) we test the dependence on $T$ and $H$ by
plotting $\frac{H^2}{T} R$ versus $t/\tau$ for $T = 1,\ 2 $ and
$H=0.5,\ 1, \ 1.5$. We found the best collapse using $\tau \sim
H^{-3}$ but the precision of our data is not high enough to distinguish between
this and the $\tau$s proposed in \cite{Villain} and \cite{Rao}.  Our
numerical results tend to confirm the $T/H^2$ dependence of $R$ even
in the early stages of the growth.

\begin{figure}
\begin{center}
\hspace{0.5cm}
  \includegraphics[width=4.75cm,angle=-90]{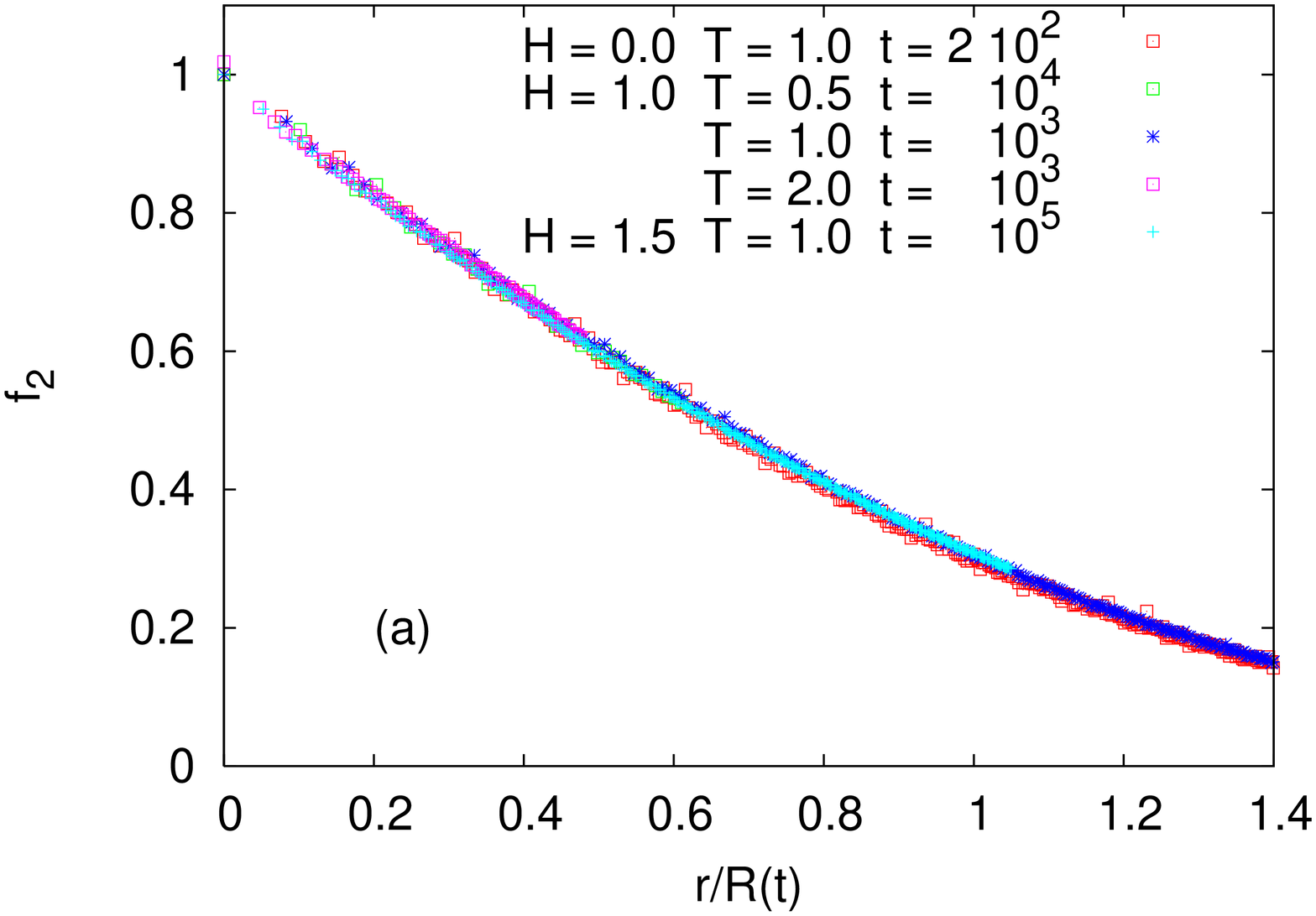}
  \includegraphics[width=4.75cm,angle=-90]{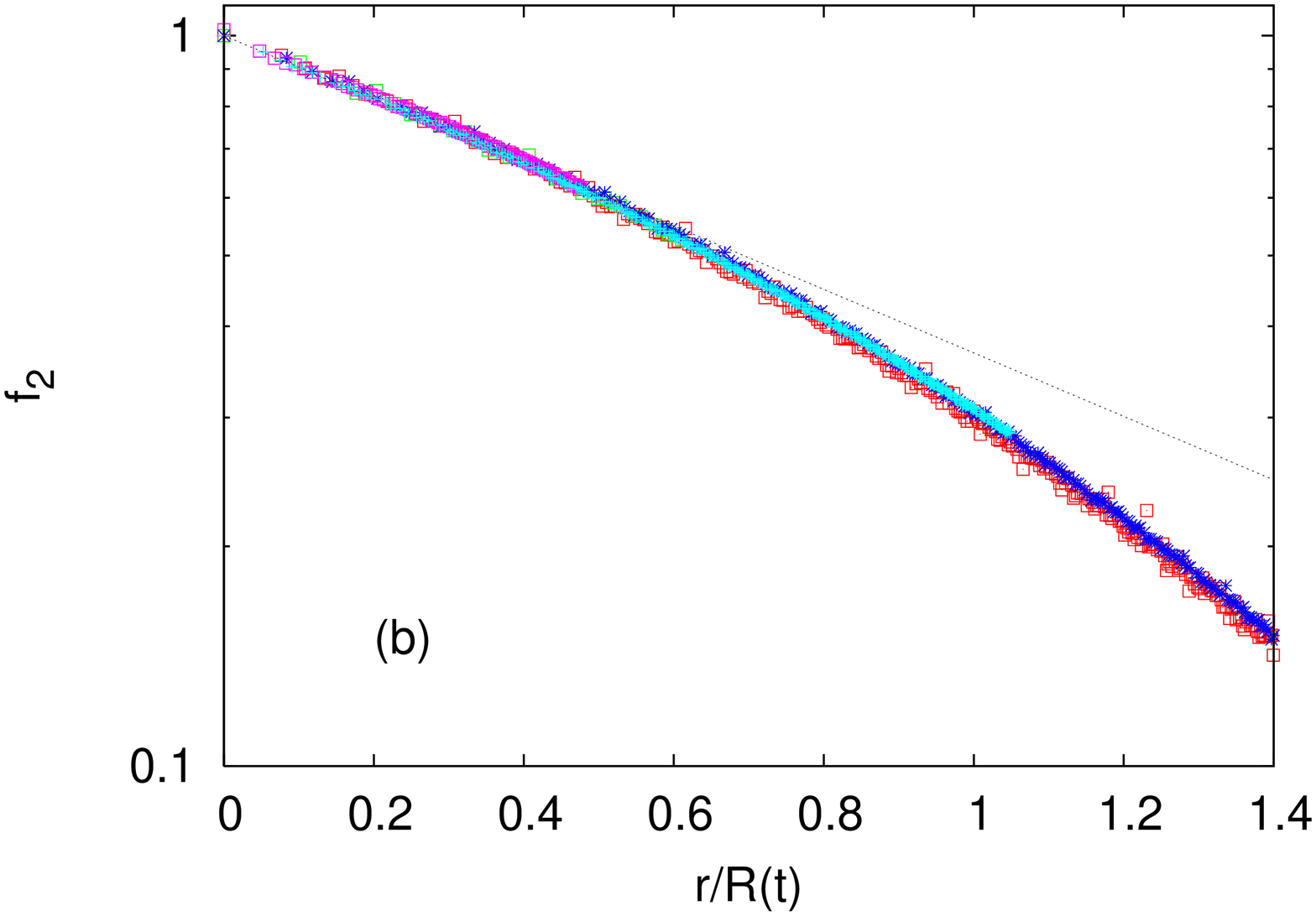}
  \caption{\footnotesize \label{fig:RF_C2_Scaling} (a)~The scaling
    function $f_2(r/R)$ for $T = 0.5, 1, 2$ and $H=1, 1.5$.
(b)~The same data in a linear-log scale showing that $f_2$ is close to an exponential at short $r/R$.}
\end{center}
\end{figure}

Since the work of \cite{Rao}, it is now clear that $f_2$ in
Eq.~(\ref{eq:C2-scaling}) is independent of $H$, and very similar to
the one of the pure system. In Fig.~\ref{fig:RF_C2_Scaling} we also
find that the scaling functions $f_2$ at different $T$ fall
on top of one another. Thus $f_2$ is independent of $H$ {\it and} $T$.

\subsubsection{The two-time self-correlation.}

It is commonly defined as
\begin{equation}
  C(t,t_w) \equiv \frac{1}{N} \sum_{i=1}^N \langle s_i(t)s_i(t_w) \rangle \,,
\end{equation}
and quantifies how two spin configurations of the same system, one
taken at $t_w$ (waiting time) and the other one at $t \geq t_w$, are
close to each other.  The angular brackets here indicate an average
over different realizations of the thermal noise. In the large $N$
limit, this quantity is self-averaging with respect to noise and
disorder induced fluctuations. This two-time function has been used as
a clock for the out of equilibrium dynamics of glassy
systems~\cite{Cuku94,Bacukupa} and we shall use this property again, in the
study of the two-time growing length and fluctuations.

The behaviour of $C$ is well understood for coarsening systems.  As
long as the domain walls have not significantly moved between $t_w$
and $t (> t_w)$ (that defines what we shall call later short time
delay), the self-correlation is given by the fluctuations of
spins that are in thermal equilibrium inside the domains. As any other
equilibrium two-time function, the self-correlation depends then only
on $t-t_w$. Later, for longer time delays, the displacement of domain
walls cannot be neglected any more and $C$ looses its
time-translational invariance. The
self-correlation can be written as a sum of two terms representing
the thermal and aging regimes:
\begin{equation}
 C(t,t_w) = C_{\rm th}(t - t_w) + C_{\rm ag}(t,t_w)
\end{equation}
with the limit conditions
\begin{eqnarray}
\begin{array}{ll}
 C_{\rm th}(0) = 1 - q_{\rm EA}\,, 
\qquad 
&
\lim_{t_w \rightarrow t^{-}}C_{\rm ag}(t,t_w) = q_{\rm EA}\,, 
\nonumber \\
 \lim_{t - t_w \rightarrow \infty}C_{\rm th}(t-t_w) = 0 \,, 
\qquad
&
\lim_{t \gg t_w}C_{\rm ag}(t, t_w) = 0 \,. \nonumber
\end{array}
\end{eqnarray}
$q_{\rm EA}$ is a measure of the order parameter and in a ferromagnetic
phase it simply equals $m^2_{eq}$, the magnetization squared.

\begin{figure}
\begin{flushright}
 \includegraphics[width=4.75cm,angle=-90]{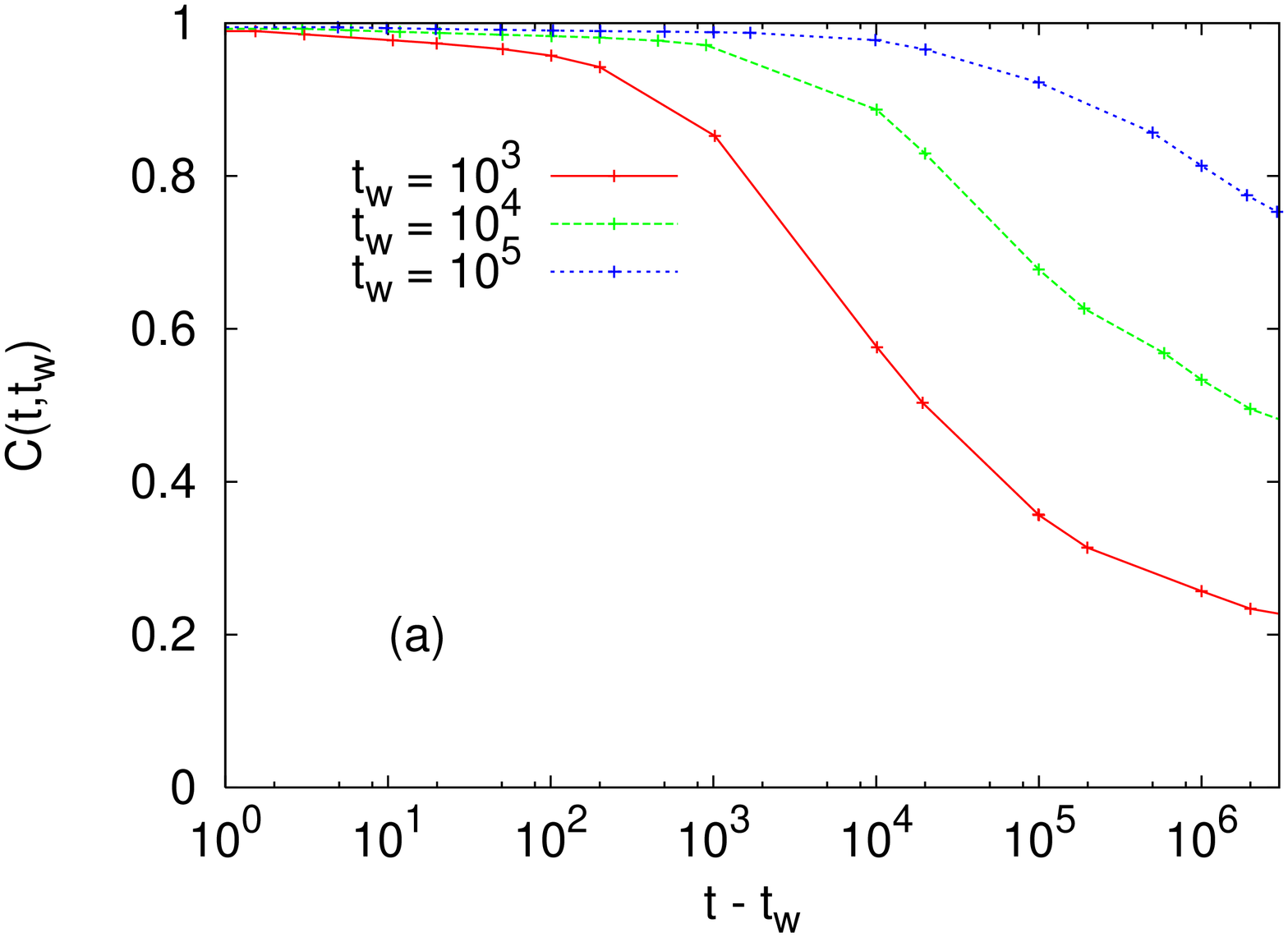}
 \includegraphics[width=4.75cm,angle=-90]{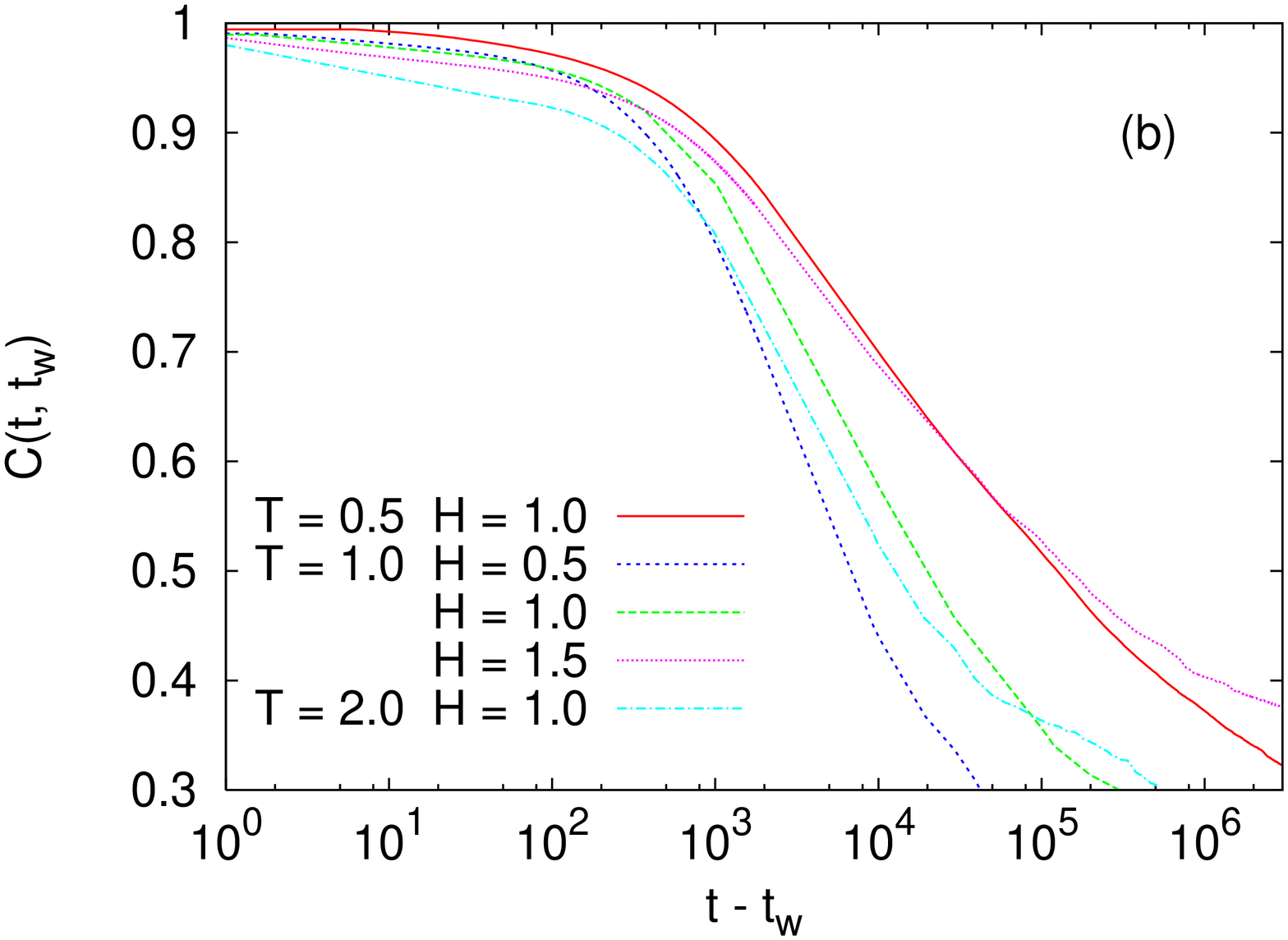}
\end{flushright}
\begin{center}
 \caption{\footnotesize \label{fig:C_Various} The global correlation
   $C$ vs $t-t_w$. (a)~$T = 1$ and $H = 1$ and different $t_w$ given
   in the key. (b)~$t_w=10^3$ at various pairs of $(T,H)$ given in the
   key.}
\end{center}
\end{figure}

In Fig.~\ref{fig:C_Various}~(a) we show the decay of the two-time
correlation $C$ as a function of the time delay $t-t_w$ for $t_w =
10^3, \, 10^4, \, 10^5$ at $T= 1$ and $H = 1$.  On each of these
curves, one can distinguish the two dynamic regimes. The longer the
waiting time the later the aging regime appears.  In
Fig.~\ref{fig:C_Various}~(b) we show the decay of the two-time
correlation as a function of time-delay for $t_w = 10^3$ and five
pairs of parameters $(T,H)$ given in the key. It is clear that the
full relaxation depends strongly on the external parameters: raising
the temperature or reducing the random field strength speeds up the
decay. For these values of $T$ and $H$, $q_{\rm EA}$ does not change much
but the decay in the aging regime does.

\begin{figure}
\begin{center}
\hspace{1cm}
  \includegraphics[width=4.6cm,angle=-90]{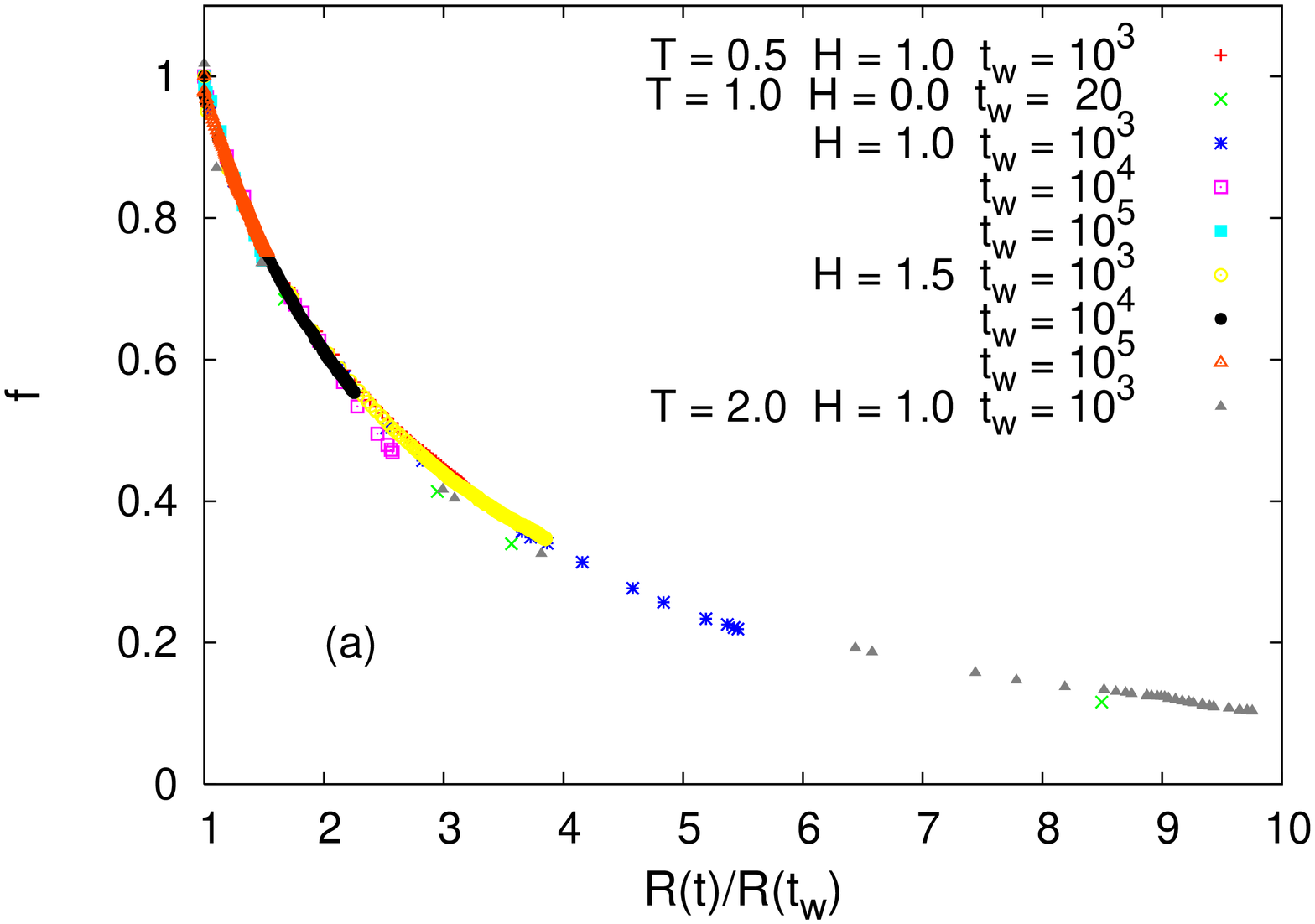}
  \includegraphics[width=4.6cm,angle=-90]{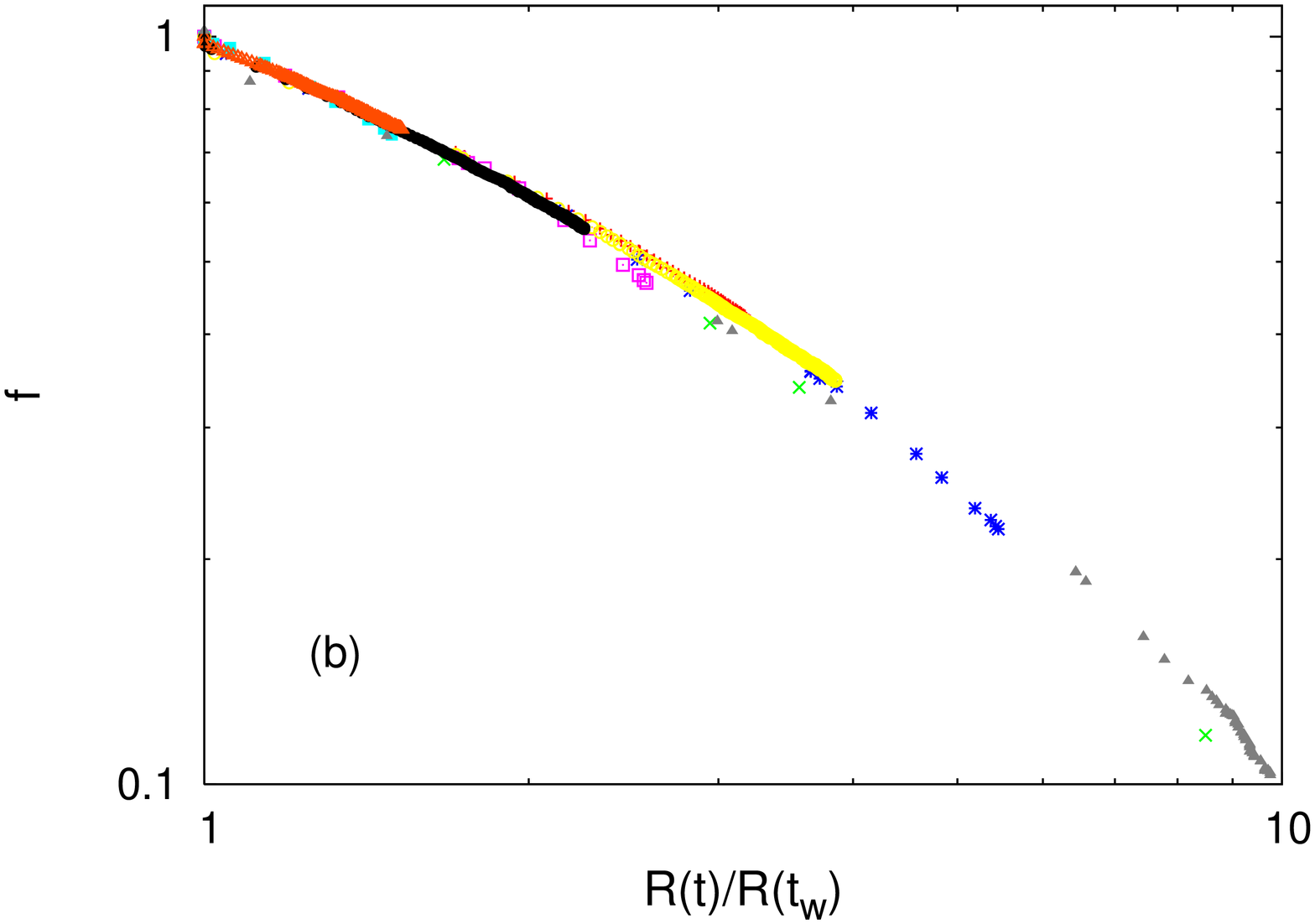}
  \caption{\footnotesize \label{fig:C_Scale} Test of the scaling and
    super-universality hypothesis. (a)~$f=C_{ag}/q_{\rm EA}$ {\it vs.}
    $R(t)/R(t_w)$ at various pairs of $(T,H)$ and $t_w$ given in the
    key. (b)~The same data in log-log scale.}
\end{center}
\end{figure}

Dynamic scaling implies that in the aging regime
\begin{equation}
 C_{ag}(t,t_w) = q_{\rm EA} \; f\left(\frac{R(t)}{R(t_w)}\right)\,,
\label{eq:superaging}
\end{equation}
with $R$ the typical length extracted from $C_2$, $f(1)=1$ and
$f(\infty)=0$. For our choice of parameters $(T,H)$, $q_{\rm EA}$ is close to unity so we can easily compute $f$ from the measured $C$ by using $f  = C_{ag}/q_{\rm EA} \simeq C/q_{\rm EA}$. Super-universality states that $f$ does not depend on $T$ and $H$. In Fig.~\ref{fig:C_Scale} we show that both hypotheses
apply to this quantity. In panel (a) we use a linear-linear scale while 
in panel~(b) we present the same data in a double logarithmic scale. 
Although the scaling function $f$ looks like a power law
it is not. One expects that its tail [$R(t)\gg R(t_w)$]
becomes a power-law with an exponent $\lambda$. The actual 
function $f$ is not known. Most of the analytic efforts in domain
growth studies are devoted to develop approximation schemes to 
derive $f$, $f_2$ and other scaling functions 
but none of them is fully successful~\cite{Alan}.

\subsubsection{The four point-correlation function.}

\begin{figure}
\begin{flushright}
 \includegraphics[width=4.75cm,angle=-90]{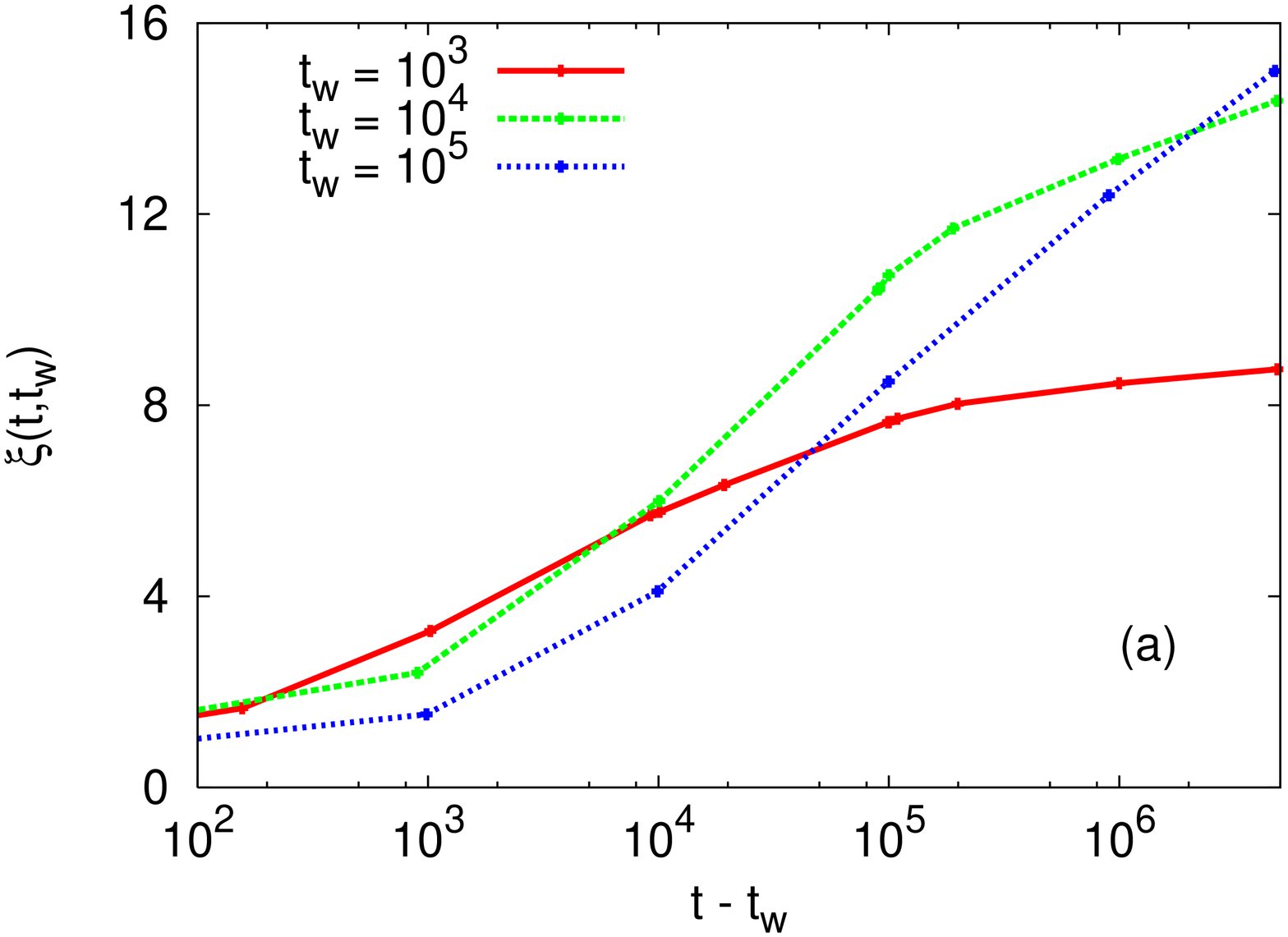}
 \includegraphics[width=4.75cm,angle=-90]{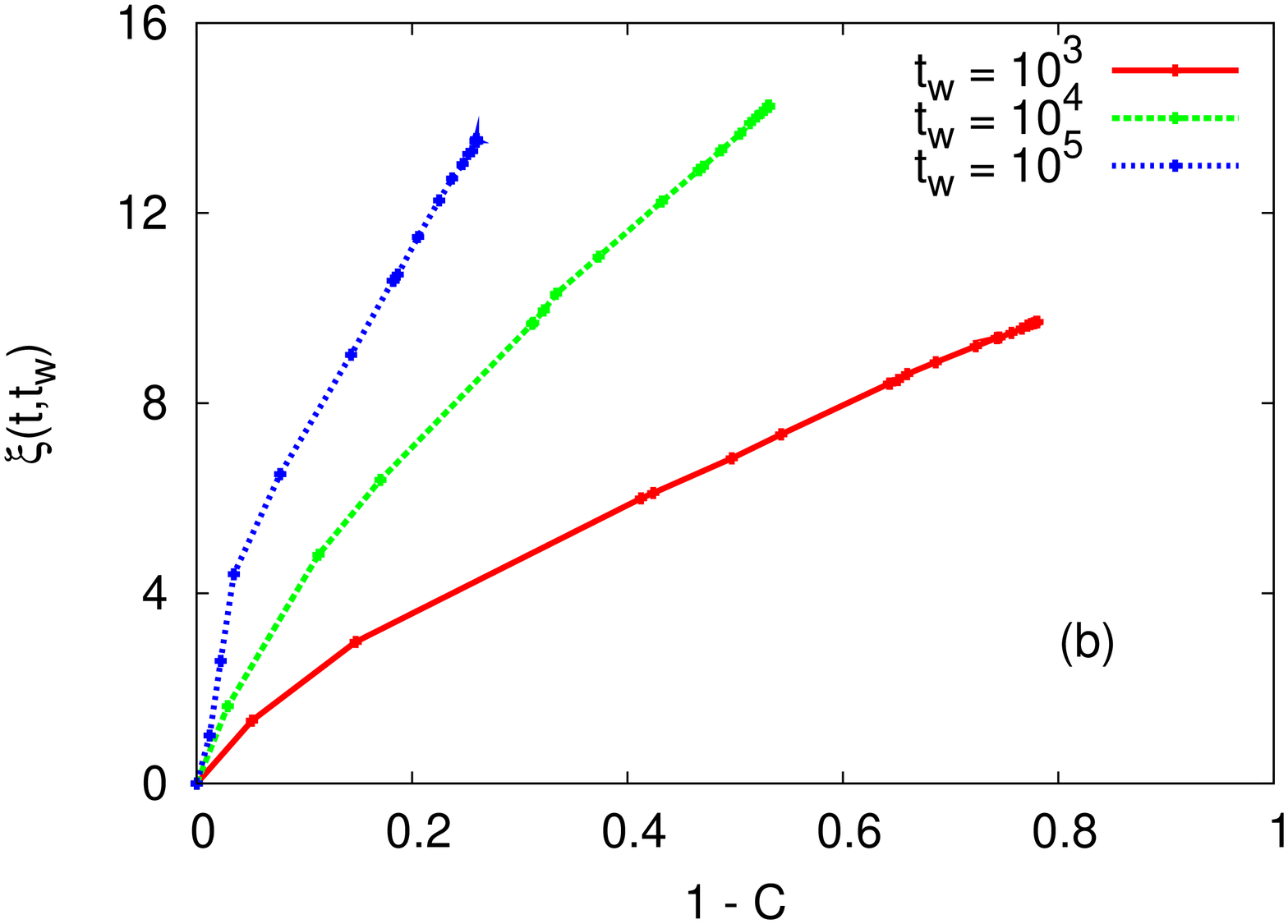}
\end{flushright}
\begin{center}
\hspace{1cm}
\includegraphics[width=4.75cm,angle=-90]{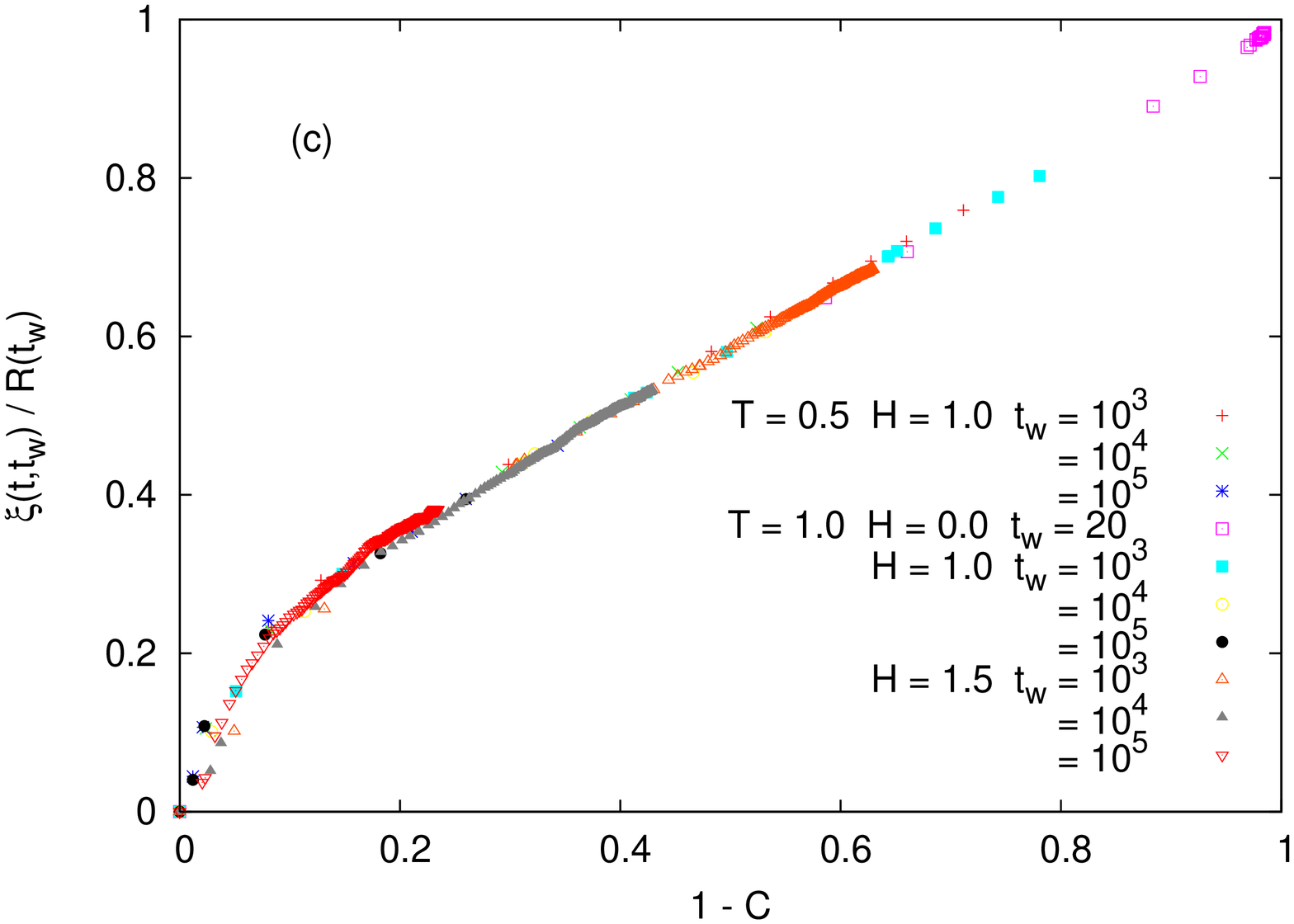}
 \caption{\footnotesize 
   \label{fig:xi} 
   The two-time correlation length, $\xi$, in the RFIM.  (a)~$\xi$ as
   a function of time-delay, $t-t_w$ for several values of $t_w$ given
   in the key at $T=1$ and $H=1$.  (b)~$\xi$ as a function of the
   global correlation in a parametric plot at $T=1$ and $H=1$.  (c)~Scaling $\xi(t,t_w) = R(t_w) \ g(C)$ at two temperatures and two values of the random field using three waiting-times $t_w$ for each
   set of parameters. The pure case $H=0, T=1$ is also included with a very short $t_w$ to avoid equilibration.}
\end{center}
\end{figure}

In order to successfully identify a growing correlation length in
glassy systems including the $3d$ EA spin-glass, one defines the
two-time two-site correlation
function~\cite{Ludovic,Castillo,Castillo-Parsaeian,Weeks,Sharon}
\begin{equation}
  C_4(r;t,t_w) \equiv 
  \langle s_i(t) s_i(t_w) s_j(t) s_j(t_w)\rangle_{\vert \vec r_i - 
\vec r_j\vert = r} \,.
\end{equation}
We extract $\xi$ from its approximate spatial exponential decay:
$C_4(r;t,t_w)-C^2(t,t_w) \propto e^{-r/\xi(t,t_w)}$ at relatively {\it short} $r/\xi$.
(Other methods, like defining the connected four spin-correlation
and extracting $\xi$ from its volume integral yield similar qualitative
results though slightly different quantitatively.)
Results of this analysis are shown in Fig.~\ref{fig:xi}~(a) where we
plot $\xi(t,t_w)$ as a function of $t$ for different $t_w$ at $T=1$
and $H=1$.  We identify a short $t-t_w$ regime that is independent of
$t_w$ (thermal regime), whereas for long $t-t_w$, time-translational
invariance is broken (aging regime). In Fig.~\ref{fig:xi}~(b) we plot
$\xi(t,t_w)$ versus $1-C(t,t_w)$ for the three same values of $t_w$,
using $t$ as a parameter. The dependence on $1-C$ and $t_w$ is
monotonic and very similar to the one obtained in the $3d$ EA
model~\cite{Ludovic} (see Fig.~\ref{fig:xi_ea}).  The thermal regime is almost
invisible here since it is contained between $C=1$ and $C=q_{\rm EA}$,
with $q_{\rm EA} \simeq 1$ for this set of parameters.  We then propose
\begin{equation}
\xi(t,t_w) = R(t_w) \; g(C)
\, . 
\label{eq:scaling-xi}
\end{equation}
The limit $g(C=1)=0$ is found by taking $t=t_w$, that corresponds to
$C=1$ [extending the scaling form (\ref{eq:scaling-xi}) to include the
thermal regime].  In this case $C_4(r; t,t)=1$. If one uses
$C_4(r;t,t) = \tilde C_4(r/\xi,C(t,t)=1)$, see
Sect.~\ref{sec:C4-super}, then $\xi(t,t)$ must vanish to obtain $C_4$
independent from $r$, and this imposes $g(1)=0$.  In the other extreme,
when $t\gg t_w$ and $C=0$ one expects $g(0)=1$.  The reason is the
following. $\lim_{t\gg t_w} C_4(r;t,t_w)=C_2(r,t)C_2(r,t_w)$, for the
temporal decoupling of $C_4$ can be done in the $t\gg t_w$
limit. Recalling that $C_2(r,t) \propto \, f_2(r/R(t))$ with
$\lim_{x\to 0} f_2(x)=1$, the only spatial contribution to
$\lim_{t\gg t_w} C_4(r;t,t_w)$ comes from the term $C_2(r,t_w) \propto
f_2(r/R(t_w))$. Using $\lim_{t\gg t_w} \xi(t,t_w) = R(t_w) g(0)$ and
further assuming that the functional forms of $C_4(x)$ and $f_2(x)$
are, to a first approximation, the same we deduce $g(0)=1$.

Figure~\ref{fig:xi}~(c), where we plot $\xi(t,t_w)/R(t_w)$ versus
$1-C(t,t_w)$ for different $t_w$, illustrates the validity of the
scaling hypothesis (\ref{eq:scaling-xi}). We see that, as expected,
$g(C=1)=0$ and it seems plausible that $\lim_{C\mapsto0}g(C) = 1$. The
scaling function $g$ is found to satisfy super-universality, \textit{i.e.}
it is independent of $H$ and $T$.

\subsubsection{$C_4$ and super-universality.}
\label{sec:C4-super}

\begin{figure}
\begin{center}
\hspace{1cm}
 \includegraphics[width=4.75cm, angle = -90]{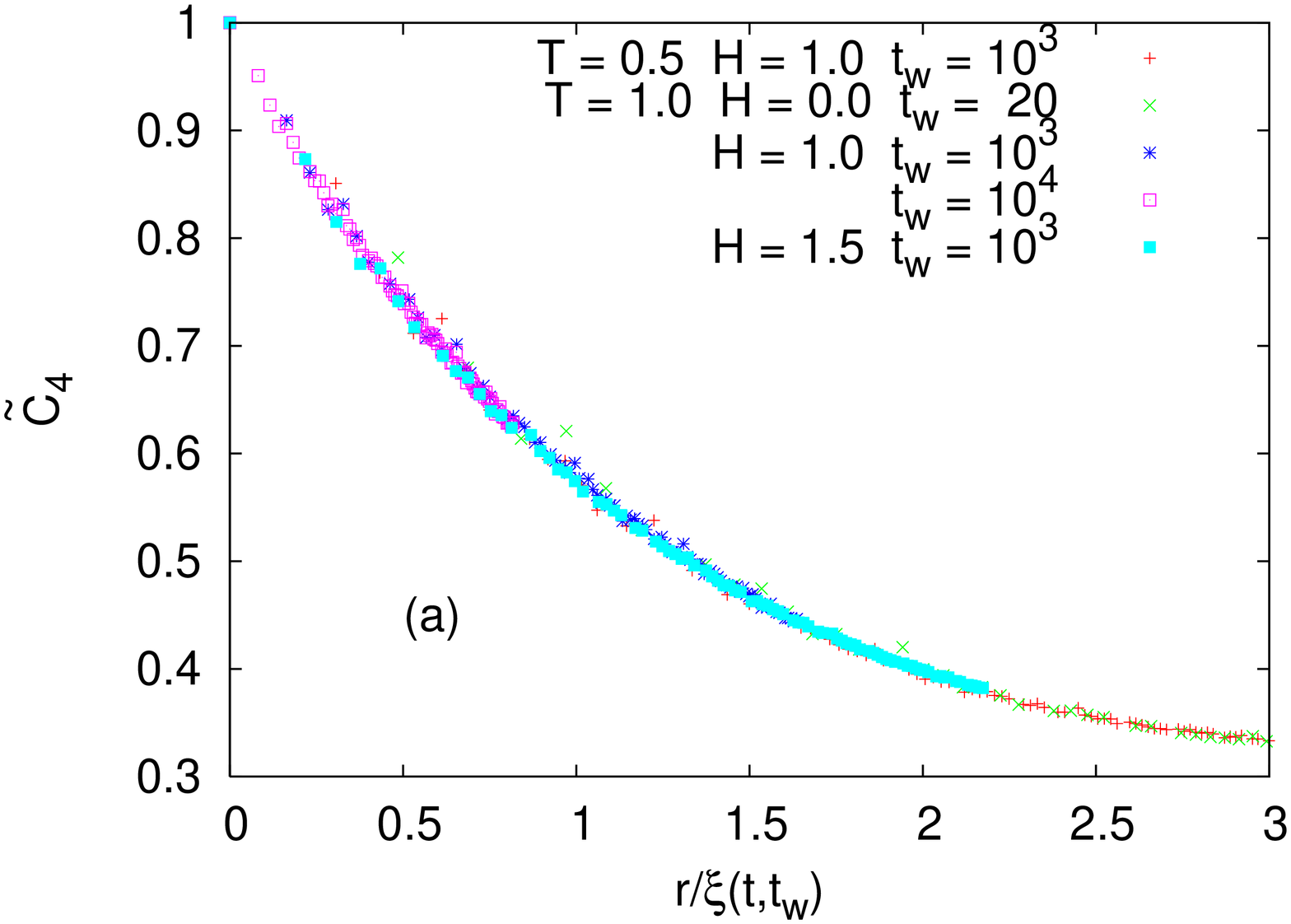}
 \includegraphics[width=4.75cm, angle = -90]{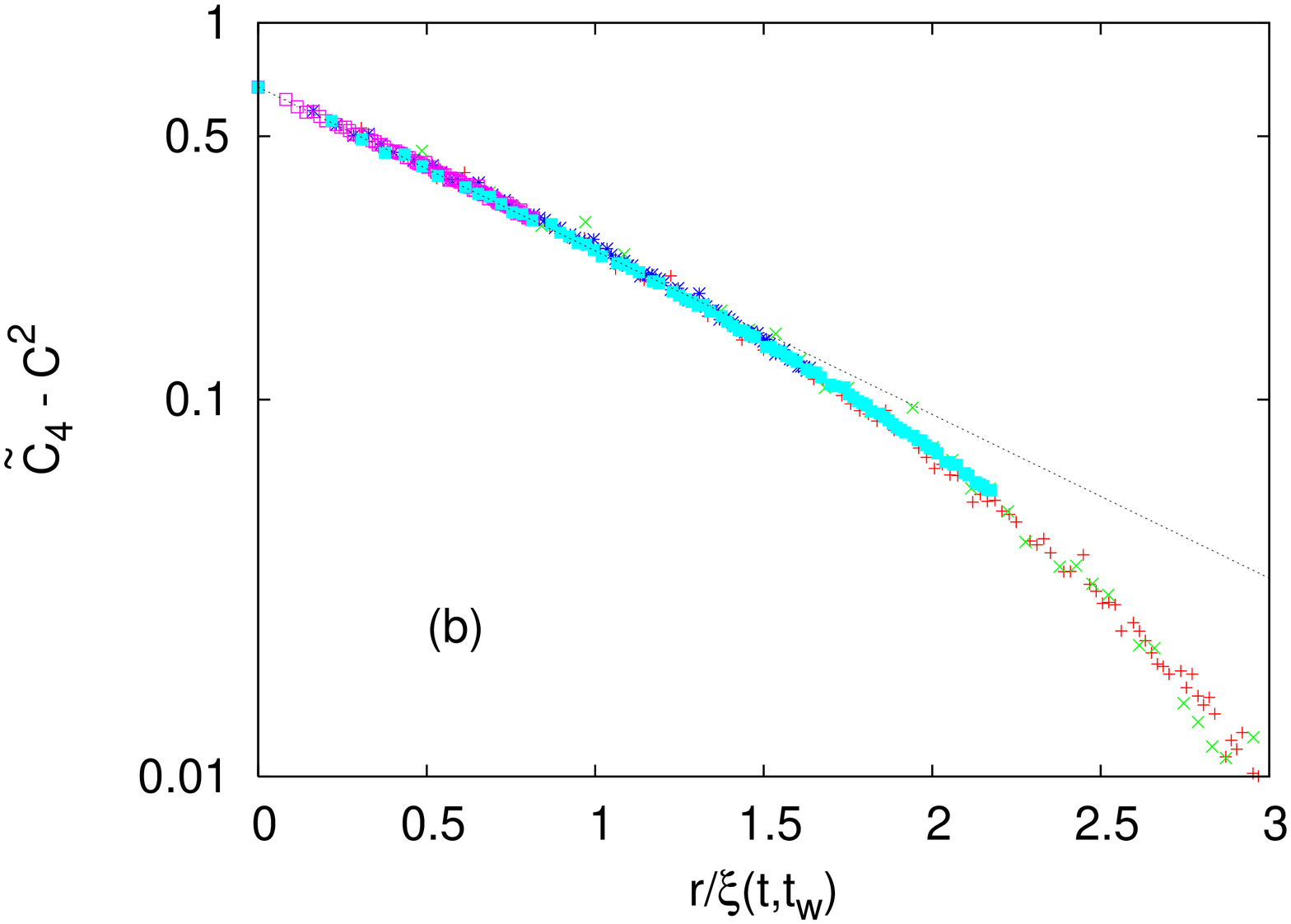}
\end{center}
\caption{\footnotesize \label{fig:C4_Various} (a)~Test of scaling, $\tilde
  C_4(r/\xi, C)$, and the super-universality of $\tilde C_4$ for the
  parameters $T$ and $H$ given in the key. Times $t$ and $t_w$ are
  chosen in such a way that $C(t,t_w) = 0.57$ in all cases. (b)~The same data in linear-log scale showing that $\tilde
  C_4-C^2$ is very close to an exponential at short $r/\xi$.}
\end{figure}

Using the monotonicity properties of $C$ as a function of $t-t_w$ and
$t_w$, and of $\xi$ as a function of $t_w$ and $1-C$ we can safely
exchange the dependence of $C_4$ on the two times by a dependence on
$\xi$ and $C$. In other words, $C_4(r,\xi,C)$ where, again for
simplicity, we did not write explicitly the dependence on $T$ and $H$. 
Now, a reasonable scaling assumption is that one can measure $r$
in units of $\xi$ such that
\begin{equation}
C_4(r,t,t_w) = \tilde C_4(r/\xi(t,t_w), C(t,t_w))
\; . 
\label{eq:C4-scaling}
\end{equation}
In Fig.~\ref{fig:C4_Various} we put this scaling form to the test and
we examine the possible super-universality of $\tilde C_4$. We
use different values of the parameters $t$, $t_w$, $T$, $H$ such that
$C=0.57$ in all cases. Both scaling and super-universality relations
are well satisfied.  Note that the scaling relation in
Eq.~(\ref{eq:C4-scaling}) can also be transformed into
\begin{equation}
   C_4(r;t,t_w) = C_4(r/R(t_w), R(t)/R(t_w))
\end{equation}
by using Eq.~(\ref{eq:superaging}). This last scaling form was also found
for the O($N$) ferromagnetic model in the large $N$ limit although the
scaling function does not have a simple exponential
relaxation~\cite{Chcuyo}.

\subsection{3d EA}
\label{subsec:3dEAXiScale}

\begin{figure}
\begin{flushright}
 \includegraphics[width=4.75cm,angle=-90]{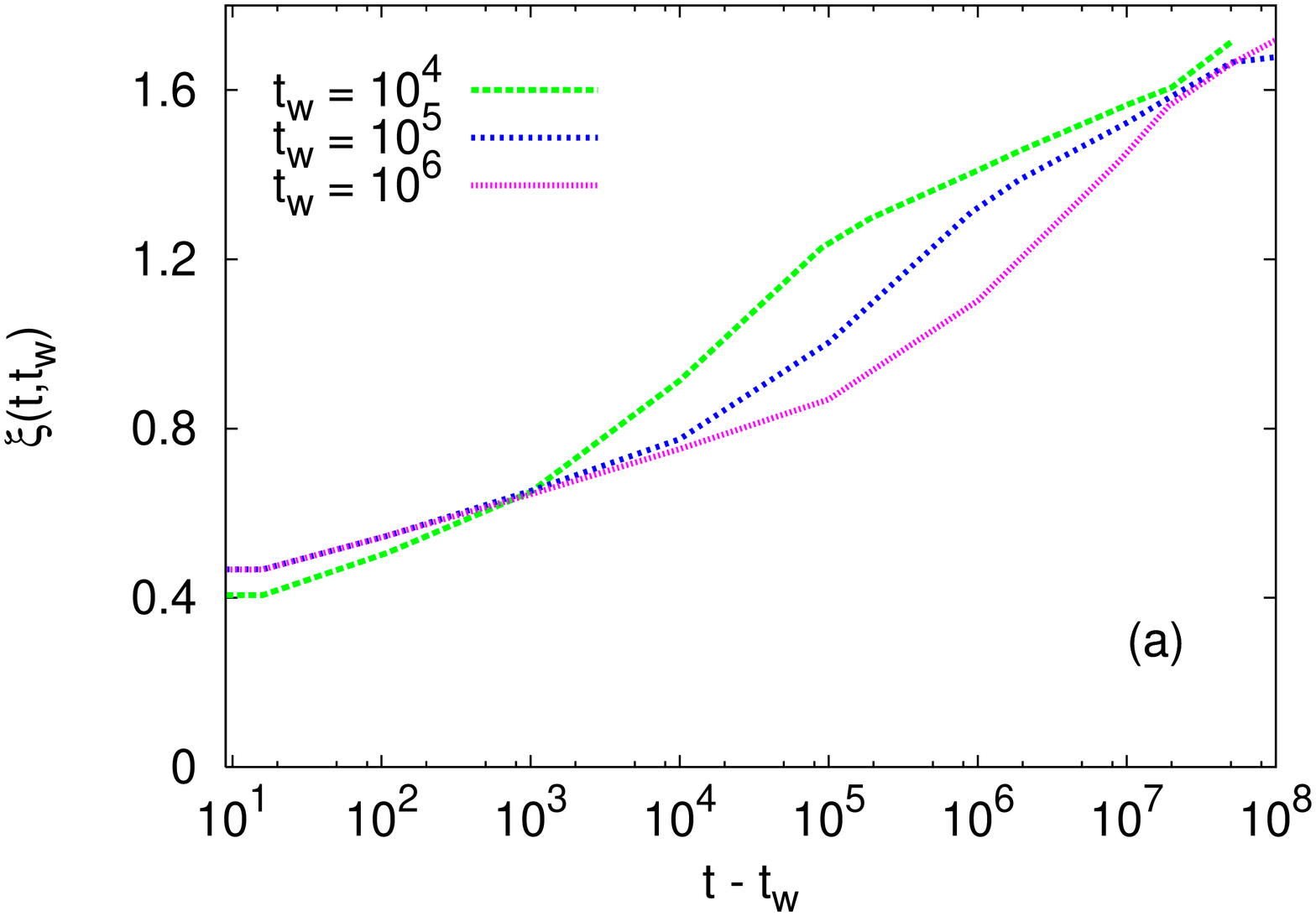}
 \includegraphics[width=4.75cm,angle=-90]{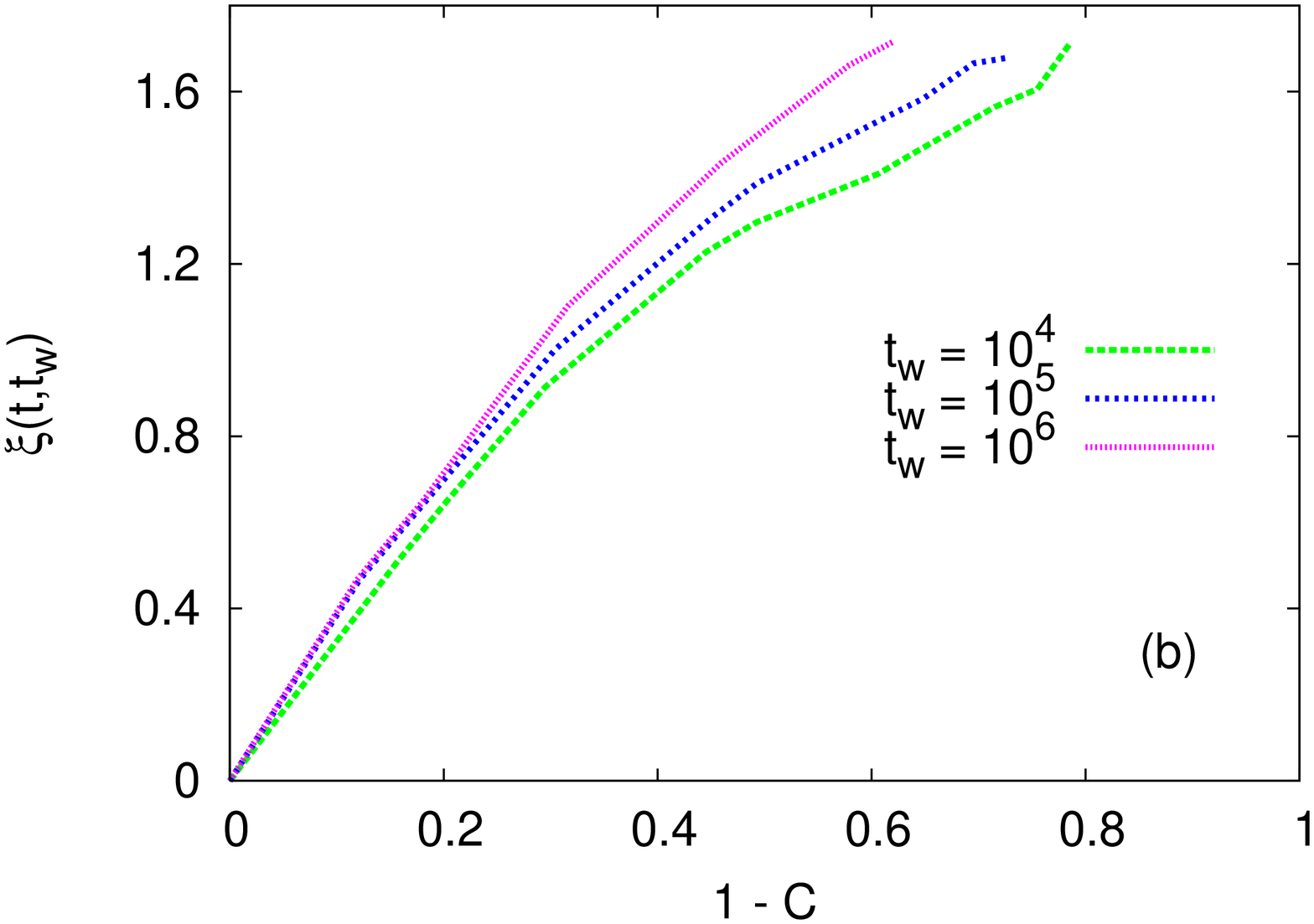}
\end{flushright}
\begin{center}
\hspace{1cm} 
\includegraphics[width=4.75cm,angle=-90]{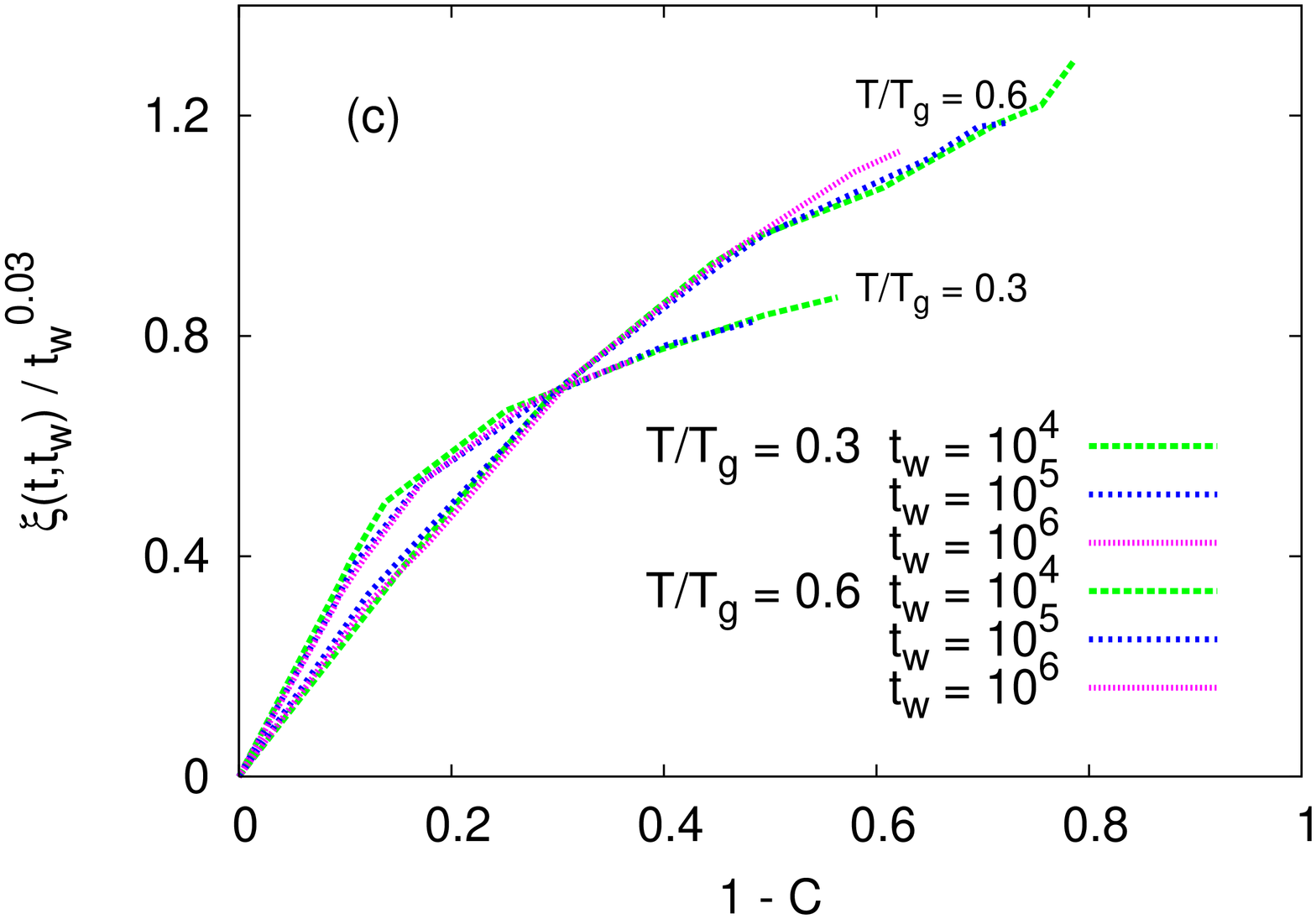}
\end{center}
\begin{center}
 \caption{\footnotesize 
   \label{fig:xi_ea}
   Study of the two-time correlation length in the $3d$ EA model.  (a)~$\xi$ as a function of time-delay $t-t_w$ for several $t_w$'s
   given in the key at $T/T_g=0.6$.  (b)~Evolution of $\xi$ with the
   global correlation in a parametric plot at $T/T_g=0.6$.  (c)~Test
   of the scaling hypothesis $\xi(t,t_w) = R(t_w) \ g(C)$ with $R(t)
   \propto t^{0.03}$ at $T/T_g = 0.3$ and $T/T_g=0.6$.}
\end{center}
\end{figure}

A detailed analysis of the relaxation properties of similar
correlations in the $3d$ EA model appeared in \cite{Ludovic}.  The
spatial one-time correlation, $C_2(r,t)$, vanishes identically in this
model due to the quenched random interactions. The two-time
self-correlation satisfies scaling with $R \sim t^{1/z(T)}$ cannot be
simply associated to a typical radius of equilibrated domains.  The
question as to whether the scaling function $f$ is super-universal is
not well posed since the $T$-dependent power $1/z(T)$ can be absorbed
in $f$.  The four-point correlation allows for the definition of a
two-time growing length scale $\xi$ that behaves qualitatively as in
Eq.~(\ref{eq:scaling-xi}).  In Fig.~\ref{fig:xi_ea} we present
$\xi(t,t_w)$ for the $3d$ EA. Its behaviour is very similar to the one
of the RFIM exposed previously, but we would like to stress the fact
that this quantity reaches much lower values in the $3d$ EA case
(around $2a$) than in the RFIM (around $15a$).
Figure~\ref{fig:xi_ea}~(c) demonstrates that the superscaling property
does not hold in the $3d$ EA model. We used $R(t) \propto t^{0.03}$
for both temperatures and the resulting $g(C)$ curves are
significantly different. It is important to remark that no
$T$-dependent power-law in $R$ would make the two curves collapse.
Turning back to the scaling of the two-time correlation and fixing the
power law, $C\propto f[(t/t_w)^{0.03}]$ one finds $f(x) \sim x^{-4.5}$
(at $T/T_g\sim 0.6$) a much faster decaying power than in the RFIM.
Note that previous estimates of the dynamic exponent using the
one-time replica overlap~\cite{3dEA-num2,3dEA-num3} yield $1/z(T=0.3
T_g) \approx 0.045$ a slightly larger value; the reason for the
discrepancy could be traced to the lack of accuracy in the determination
of $\xi$ and then $R$.

\subsection{Colloidal glasses}

The structure factor of colloidal suspensions and Lennard-Jones
mixtures are obviously very different from the one of a
sample undergoing ferromagnetic ordering. Still, two-time self-correlations
satisfy scaling with $R(t) \propto t^{1/z}$ although a clear
interpretation of $R$ is not available.

Castillo and Parsaeian studied $\xi$ in a Lennard-Jones mixture of
particles undergoing a glassy arrest. One notices that, at short time delays
($t-t_w\sim 10$ molecular dynamic units), $\xi$ is monotonic with
respect to $t-t_w$ {\it and} $t_w$ in this system, while one needs to
reach much longer time delays (and indeed go beyond the simulation
window) in the $3d$ EA and RFIM cases [{\it cfr.}
Figs.~\ref{fig:xi}~(a) and \ref{fig:xi_ea}~(a) to the first panel in
Fig.~2 in \cite{Castillo-Parsaeian}]. A form like (\ref{eq:scaling-xi}) 
describes $\xi$ in this case too with $R(t) \sim t^{1/z}$ and 
$1/z\sim 0.1$. 

The two-time correlation length of colloidal suspensions was analysed
in \cite{Weeks} using a mapping to a spin problem. The data for $\xi$
remains, though, quite noisy and although a similar trend in time
emerges the precise functional form is hard to extract.

\subsection{Summary}

In short, the macroscopic correlations in all these systems admit the
same dynamic scaling analysis although there is no clear
interpretation of $R$ as a domain size in the case of the $3d$ EA and
colloidal suspensions.
  
\section{Fluctuations}
\label{sec:fluctuations}

An approach apt to describe problems with and without
quenched randomness focuses on thermally induced
fluctuations~\cite{Chamon-Cugliandolo}. The local dynamics can then
be examined by studying two-time spin-spin functions which, instead of
being spatially averaged over the whole bulk, are only averaged over a
coarse-graining cell with volume $V_r = (2l)^3$ centered at some site
$r$~\cite{Castillo}:
\begin{equation}\label{eq:Cr}
	C_r(t,t_w) \equiv \frac{1}{V_r} \sum_{{\overrightarrow{r_i} \in V_r}}
s_i(t) s_i(t_w) 
\; . 
\end{equation}
One can then characterize the fluctuations by studying their
probability distribution function (pdf) $\rho(C_r;t,t_w,l,L,T,H)$ with
mean value $C(t,t_w)$.

In general, 
the variation of $\rho(C_r)$ with the size of the coarse-graining
boxes is as follows.  For $l < R$ the pdf is peaked around $q_{\rm
  EA}$ and has a fat tail towards small values of $C_r$ including
negative ones.  Indeed, well in the coarsening regime, most of the
small coarse-grained cells fall inside domains and one then expects to
find mostly a thermal equilibrium distribution -- apart from the
tail. For larger values of $l$ such as $l \simeq R$, a second peak
close to $C$ appears and the one at $q_{\rm EA}$ progressively diminishes
in height. For still larger values of $l$, the peak at $q_{\rm EA}$
disappears and a single peak centered at $C$ (the mean value of the
distribution) takes all the pdf weight.

\begin{figure}
\begin{flushright}
 \includegraphics[width=4.75cm,angle=-90]{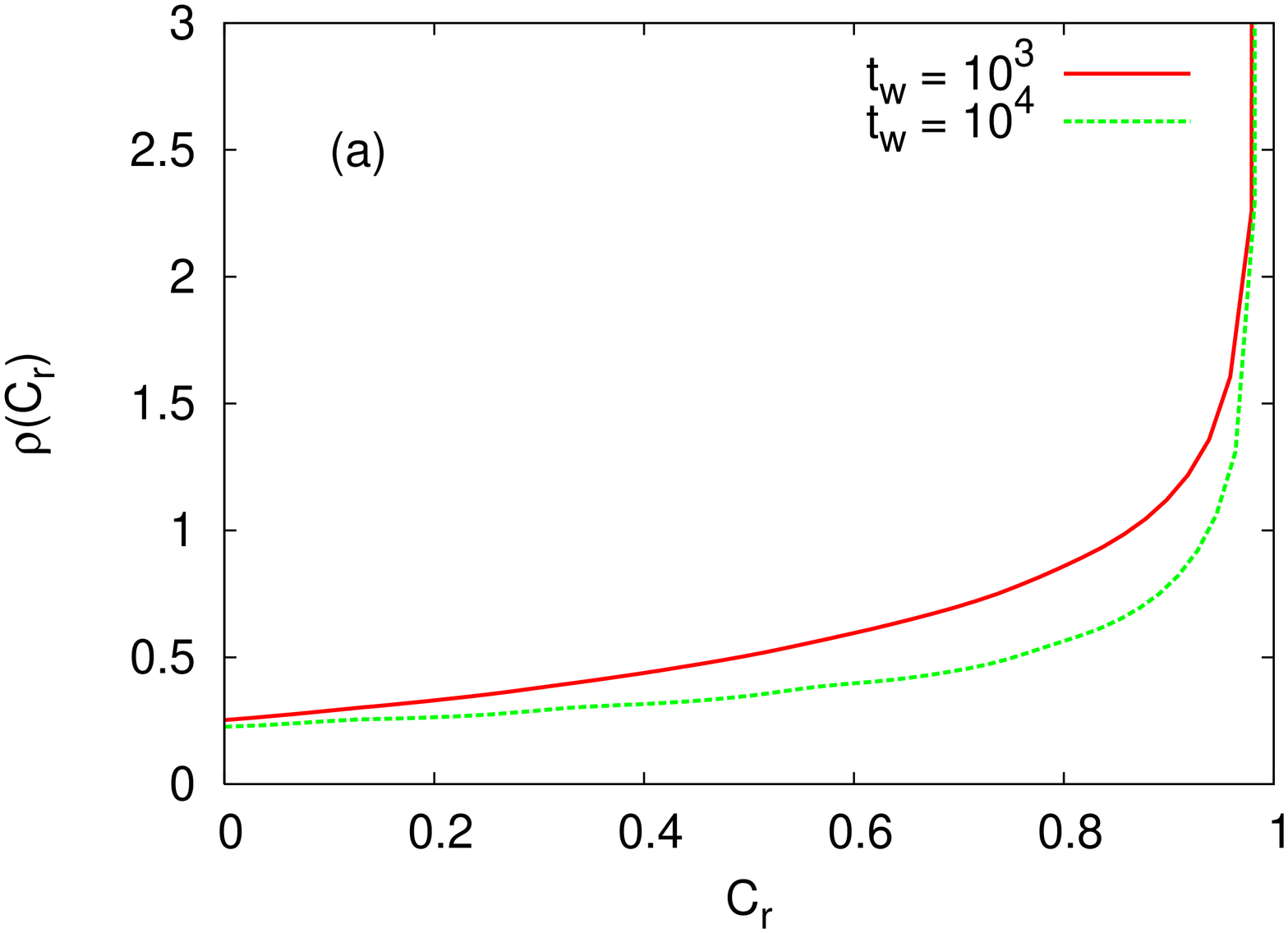}
 \includegraphics[width=4.75cm,angle=-90]{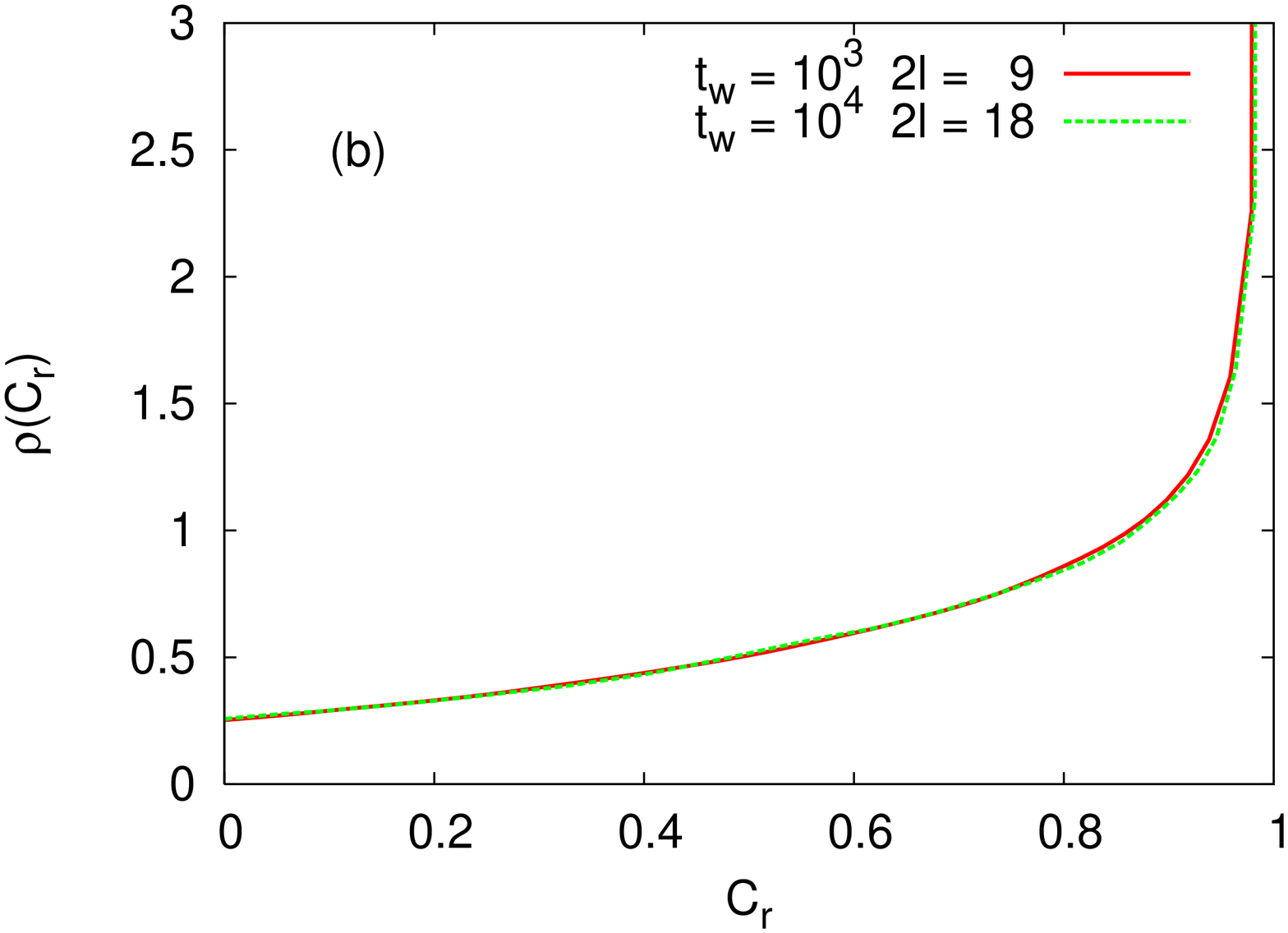}
\end{flushright}
\begin{center}
  \caption{\footnotesize \label{fig:RF_pdf_C} Pdf of local two-time
    functions $C_r$ in the RFIM at $T=1$ and $H=1$. The waiting-times
    are given in the key and time $t$ is chosen such that
    $C(t,t_w)=0.6$. (a)~$C_r$ is coarse-grained on boxes of linear
    size $l=9$. (b)~$C_r$ is coarse-grained on boxes with variable
    length $l$ so as to keep $l/\xi(t,t_w) \simeq 0.7$ constant.  The
    collapse is much improved with respect to panel (a).}
\end{center}
\end{figure}

\begin{figure}
\begin{flushright}
 \includegraphics[width=4.75cm,angle=-90]{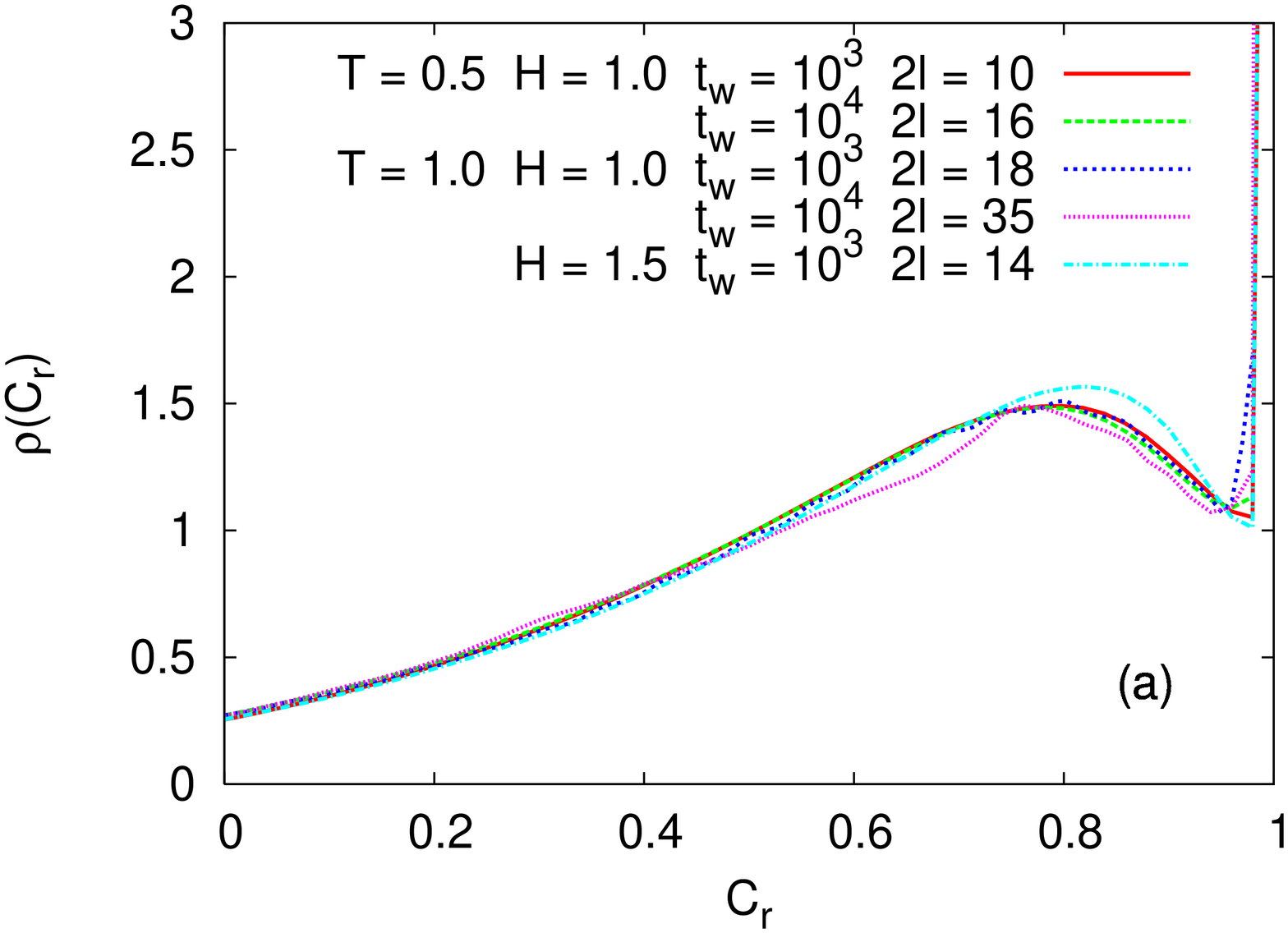}
 \includegraphics[width=4.75cm,angle=-90]{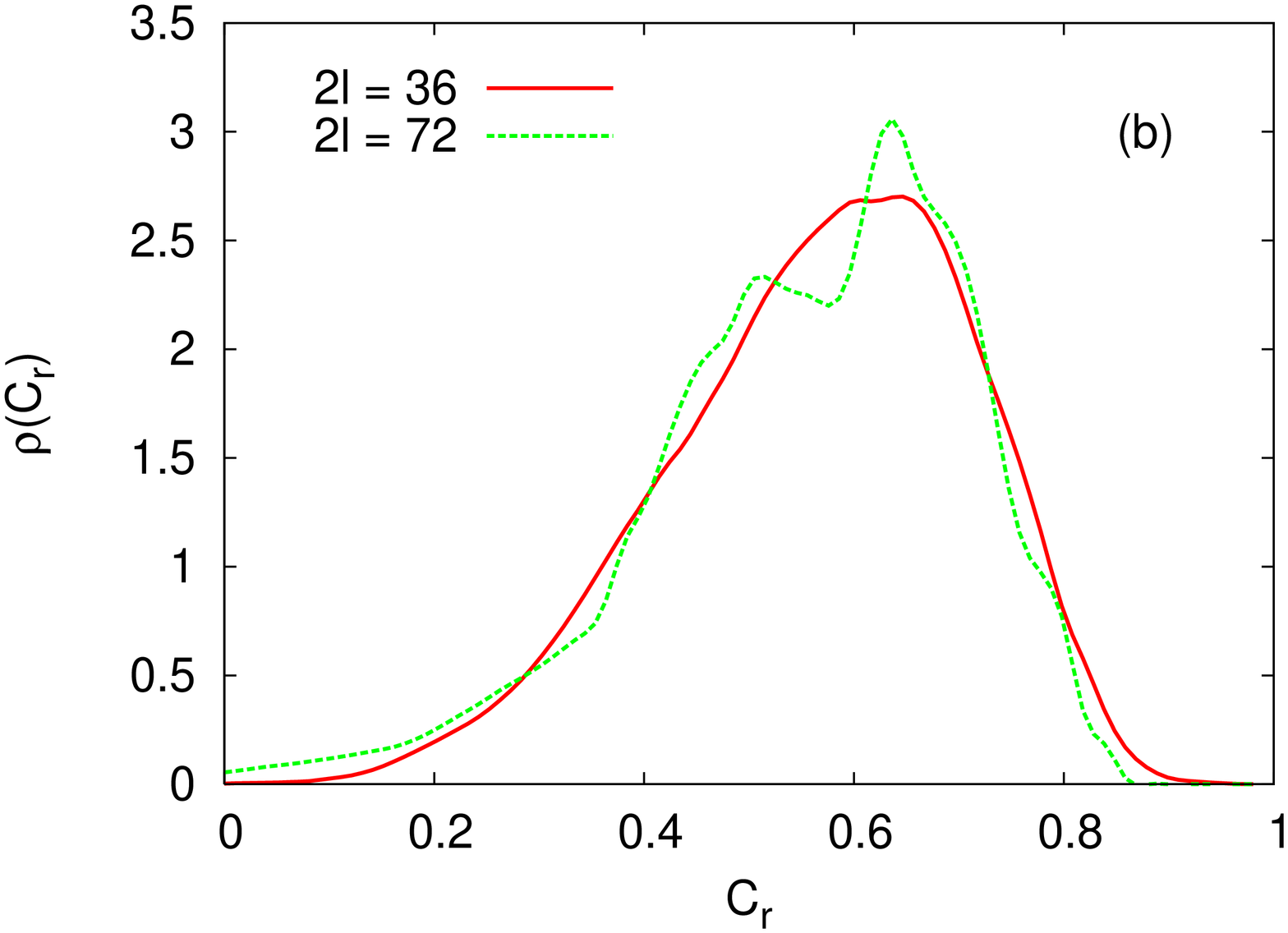}
\end{flushright}
\begin{center}
  \caption{\footnotesize \label{fig:RF_pdf_Cbis} Test of the scaling
    and super-scaling hypothesis. The two pairs of $t$ and $t_w$ are
the same as in Fig.~\ref{fig:RF_pdf_C} and $C=0.6$ as well. (a)~$l/\xi \simeq 1.4$. (b)~$l/\xi \simeq 2.9$. }
\end{center}
\end{figure}

At fixed temperature and field, the pdf $\rho(C_r;t,t_w,l,L)$ in the
RFIM depends on four parameters, two times $t$ and $t_w$ and two
lengths $l$ and $L$. In the {\it aging} regime the dependence on $t$
and $t_w$ can be replaced by a dependence on $C(t,t_w)$ and
$\xi(t,t_w)$, the former being the global correlation and the latter
the two-time dependent correlation length. Indeed, $C(t,t_w)$ is a
monotonic function on the two times [{\it cfr.}
Fig.~\ref{fig:C_Various}~(a)] and $\xi$ is a growing function of $t$
({\it cfr.} Fig.~\ref{fig:RF_R}), thus allowing for the inversion
$(t,t_w) \rightarrow (C,\xi)$.  Note that we do not need to enter the
aging, coarsening regime to propose this form.
%$\xi$ is defined even for short $t-t_w$, that is to say in the 
%thermal regime, where it is also monotonic with respect to $t_w$ and $C$. 
One can now make the natural scaling assumption that the pdfs depend
on $\xi$, the coarse-graining length $l$, and the system linear size $L$
through the ratios $l/\xi$ and $l/L$. In the end, the pdfs
characterizing the heterogeneous aging of the system read
\begin{equation}
\label{eq:rho_Cr_scaling}
	\rho(C_r;C(t,t_w),l/\xi(t,t_w),l/L)
\; .
\label{eq:scaling-pdf}
\end{equation}
We numerically test this proposal by assuming that the thermodynamic
limit applies and the last scaling ratio vanishes identically.
Figure~\ref{fig:RF_pdf_C}~(a) shows the pdfs at two pairs of times $t$
and $t_w$ such that the global correlation $C(t,t_w)$ is the same, and
$l=9$. It is clear that the two distributions are different. In panel
(b)~we further choose $l$ so that $l/\xi \simeq 0.7$ is also
fixed. The two distributions now collapse as expected from the scaling
hypothesis Eq.~(\ref{eq:scaling-pdf}).  Note that another peak at
$C=-1$ exists, though with a lower weight.
Figure~\ref{fig:RF_pdf_Cbis}~(a) and (b) show the scaling for $l/\xi
\simeq 1.4$ and $l/\xi \simeq 2.9$, respectively.  While the collapse
is still good in the case of panel~(a), it is not satisfactory in
panel~(b). Indeed, this plot suffers from the fact that the
thermodynamic limit is far from being reached ($l/L\sim 0.15$ is not
so small).

%A real space analysis shows that local regions
%exhibiting a $C_r \sim q_{\rm EA}$ (that is to say at thermal
%equilibrium) are inside spins up (resp. spins down) domains.

In Fig.~\ref{fig:RF_pdf_Cbis}~(a) we used several values of $T$ and $H$
and we found that all pdfs collapse on the
same master curve.  We conclude that as long as coarse-graining
lengths are not too close to the system size, the pdf of local
correlation satisfy the scaling (\ref{eq:rho_Cr_scaling}) with a
scaling function that is super-universal.

Let us now compare the forms of the pdfs in the RFIM and $3d$ EA
model. In the RFIM the peak at $q_{\rm EA}$ is visible until $l/\xi
\simeq 2$. Given that in this model $\xi$ is quickly rather large
($\xi$ reaches $15a$ in the simulation time-window) one has a
relatively large interval of $l$ for which the peak at $q_{\rm EA}$
can be easily seen. Instead, in the $3d$ EA the two-time correlation
length grows very slowly and reaches only $\xi \sim 2a$ in similar
times, meaning that the peak at $q_{\rm EA}$ is hardly visible as soon
as one coarse-grains the two-time observables~\cite{Ludovic}.

\begin{figure}
\centerline{
 \includegraphics[width=4.75cm,angle=-90]{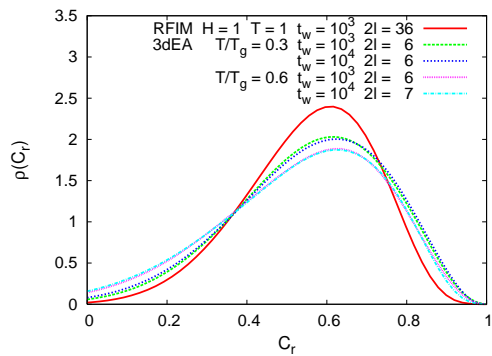}
}
\begin{center}
 \caption{\footnotesize \label{fig:EA-pdf} 
Pdf of local correlations in the $3d$ EA model at $T/T_g = 0.3$ and 
$T/T_g=0.6$, for two waiting-times $t_w$ such that 
$C=0.6$ and $l/\xi=2.9$. The solid line (red)
displays the super-universal pdf in the RFIM.}
\end{center}
\end{figure}

Figure~\ref{fig:EA-pdf} demonstrates that the pdf of local
correlations is not super-universal with respect to $T$ in the $3d$ EA
model, and compares the functional form at two temperatures,
$T/T_g=0.3$ and $T/T_g= 0.6$, with the one in the RFIM. The global
correlation, $C$, and the ratio of coarse-graining to correlation
lengths, $l/\xi$, are the same in all curves.  Although qualitatively
similar, the pdf in the RFIM and $3d$ EA models are different, with
the RFIM one being more centered around the global value.

The study of Lennard-Jones mixtures in~\cite{Parsaeian-Castillo} used
a constant coarse-graining length and the pdfs of local correlations
at constant $C$ showed a slow drift that should be cured by taking
into account the variation of $\xi$. In colloidal suspensions the
scaling form (\ref{eq:scaling-pdf}) is well satisfied~\cite{Weeks}.
In the context of coarsening phenomena these pdfs are to be compared
to the ones calculated for the O($N$) model in its large $N$
limit~\cite{Chcuyo}.

\section{Conclusion}
\label{sec:conclusions}

We performed an extensive analysis of the dynamics of the RFIM in its
coarsening regime.  We showed that the equal-time correlation
functions, global two-time correlation functions, and the four point
correlation functions obey scaling and super-universality relations in
the aging regime. The scaling relations, by means of the typical
growing length, $R \propto \ln t/\tau$, reveal a non-trivial
time-invariance for these statistical objects. Super-universality
encodes the irrelevance of quenched randomness and temperature on the
scaling functions and it is demonstrated by the 
fact that they are the same as for the pure Ising case.

In the $3d$ EA, similar scaling forms were found for global two-time
correlations and four-point correlations~\cite{Ludovic}.  The function
$R(t)$ could be associated to a domain radius though a clearcut
confirmation of this is lacking. On the contrary, the results of
recent large scale simulations have been interpreted as evidence for
an SK-like dynamic scenario~\cite{Victor}.  The one-time function
playing the role of the domain radius is a very weak power law,
$t^{0.03}$ at $T/T_g\sim 0.3-0.6$, and, in consequence, the two-time
correlation length reaches much shorter values than in the RFIM in
equivalent simulation times.  Super-universality (with respect to
temperature) does not apply in this case.

A similar scenario applies to the Lennard-Jones
mixtures~\cite{Parsaeian-Castillo} and colloidal
suspensions~\cite{Weeks}. The two-time correlation length remains also
very short in accessible numerical and experimental times.

In all these systems the analysis of local fluctuations of two-time
functions leads to scaling of their probability distribution
functions. In the RFIM these also verify super-scaling with respect to 
$T$ and $H$.  In the $3d$
EA they do not. The intriguing possibility of a kind of super-scaling
in colloidal suspensions (with respect to concentration) has been
signaled in \cite{Weeks} and deserves a more careful study.

We conclude that all these systems, with {\it a priori}
very different microscopic dynamic processes admit a similar 
dynamic scaling descriprion of their macroscopic and mesoscopic
out of equilibrium evolution. 

\ack We thank L. D. C. Jaubert and A. Sicilia for very useful
discussions and T. Malakis and V. Mart\'{i}n-Mayor for helpful
correspondence.  LFC is a member of Institut Universitaire de France.

\end{document}